\documentclass[
     reprint,
     twocolumn,
    superscriptaddress,
    amsmath, amssymb,
    aps,
    prx,
    floatfix,
]{revtex4-2}

\usepackage{newtxtext, newtxmath}
\usepackage{color, xcolor}
    \definecolor{mygreen}{rgb}{0,0.6,0}
    \definecolor{mygray}{rgb}{0.5,0.5,0.5}
    \definecolor{mymauve}{rgb}{0.58,0,0.82}
    \definecolor{myyellow}{rgb}{0.9, 1, 1}
    \definecolor{myblue1}{RGB}{64,57,144}
    \definecolor{myblue2}{RGB}{128,166,226}
    \definecolor{myblue3}{RGB}{70,217,255}
    \definecolor{myyellow}{RGB}{251,221,133}
    \definecolor{myred1}{RGB}{207,67,62}
    \definecolor{myred2}{RGB}{244,111,67}

    \definecolor{dgreen}{rgb}{0.0,0.545,0.0}
\usepackage{amsmath, mathtools, commath, bm, amssymb}%mathrsfs, 
\usepackage{framed}
\usepackage{extarrows}
\usepackage{booktabs}
\usepackage{tabularx, multirow}

\usepackage[unicode=true, bookmarks=true, bookmarksnumbered=false, bookmarksopen=false, breaklinks=false, pdfborder={0 0 1}, backref=false, colorlinks=false, hidelinks]{hyperref}
    \hypersetup{linkcolor=black, urlcolor=black, citecolor=black, pdfstartview={FitH}}

\usepackage{amsthm} % 现在加载 amsthm 就不会报错了

\newtheoremstyle{normaltheorem}
  {3pt}   % Space above
  {3pt}   % Space below
  {\normalfont} % Body font
  {}      % Indent amount
  {\bfseries} % Theorem head font
  {.}     % Punctuation after theorem head
  { }     % Space after theorem head
  {}      % Theorem head spec

\theoremstyle{normaltheorem}
\newtheorem{theorem}{Theorem}

\usepackage{algorithm}
\usepackage{algpseudocode}

\usepackage{graphicx}
\usepackage{geometry}
\usepackage{graphics}
    \graphicspath{{./figures/}}
\usepackage{setspace}

\DeclareSymbolFont{CMlargesymbols}{OMX}{cmex}{m}{n}
\let\sumop\relax\let\prodop\relax
\DeclareMathSymbol{\sumop}{\mathop}{CMlargesymbols}{"50}
\DeclareMathSymbol{\prodop}{\mathop}{CMlargesymbols}{"51}

\SetSymbolFont{operators}{normal}{OT1}{ntxtlf}{m}{n}
\SetSymbolFont{operators}{bold}{OT1}{ntxtlf}{b}{n}

\usepackage[braket, qm]{qcircuit}

\renewcommand{\vec}[1]{{\bm{#1}}}

\newcommand{\bn}{\vec{\nabla}}
\renewcommand{\ge}{\geqslant}
\renewcommand{\le}{\leqslant}
\newcommand{\ee}{\mathrm{e}}

\renewcommand{\ket}[1]{|#1\rangle}

\newcommand{\T}{^\mathrm{T}}

\newcommand{\mc}{\mathcal}
\newcommand{\mr}{\mathrm}
\newcommand{\mbb}{\mathbb}

\newcommand{\EQ}{\begin{equation}}
\newcommand{\EN}{\end{equation}}
\newcommand{\EQA}{\begin{eqnarray}}
\newcommand{\ENA}{\end{eqnarray}}

\newcommand{\lrr}[1]{\left(#1\right)}
\newcommand{\lrs}[1]{\left[#1\right]}
\newcommand{\lrc}[1]{\left\{#1\right\}}

\newcommand{\lrn}[1]{\left\vert#1\right\vert}
\newcommand{\lrN}[1]{\left\Vert#1\right\Vert}

\makeatletter
\newcommand{\currentfontsize}{\number\f@size pt}
\makeatother

\newcommand\ii{\mathrm{i}}

\newcommand{\firtitle}[1]{\vspace{1em} \noindent \textbf{\large #1} \newline \noindent}
\newcommand{\sndtitle}[1]{\noindent \textbf{#1} \newline \noindent}

\newcommand{\beginsupplement}{%
	\setcounter{table}{0}
	\renewcommand{\thetable}{S\arabic{table}}%
	\setcounter{figure}{0}
	\renewcommand{\thefigure}{S\arabic{figure}}%
	\setcounter{equation}{0}
	\renewcommand{\theequation}{S\arabic{equation}}%
	\setcounter{section}{0}
	\renewcommand{\thesection}{\arabic{section}}%
}

\newtheorem{Q}{Question}

\usepackage{nomencl}
\makenomenclature
\begin{document}

\title{Quantum simulation of real-world nonlinear dynamics via Koopman method}

\author{Baoyang Zhang}
\affiliation{State Key Laboratory for Turbulence and Complex Systems, School of Mechanics and Engineering Science, Peking University, Beijing 100871, China}

\author{Dong An}
\affiliation{Beijing International Center for Mathematical Research, Peking University, Beijing 100871, China}

\author{Zhaoyuan Meng}
\affiliation{Institute of Mechanics, State Key Laboratory of Nonlinear Mechanics, Chinese Academy of Sciences, Beijing 100190, China}

\author{Yefei Yu}
\affiliation{Beijing Academy of Quantum Information Sciences, Beijing 100193, China}

\author{Xiaoxiao Xiao}
\affiliation{Beijing Academy of Quantum Information Sciences, Beijing 100193, China}

\author{Zhen Lu}
\email{zhen.lu@pku.edu.cn}
\affiliation{State Key Laboratory for Turbulence and Complex Systems, School of Mechanics and Engineering Science, Peking University, Beijing 100871, China}

\author{Yue Yang}
\email{yyg@pku.edu.cn}
\affiliation{State Key Laboratory for Turbulence and Complex Systems, School of Mechanics and Engineering Science, Peking University, Beijing 100871, China}
\affiliation{HEDPS-CAPT, Peking University, Beijing 100871, China}

\date{\today}

\begin{abstract}
Nonlinear dynamics is ubiquitous in nature, ranging from chemical pattern formation to ocean circulation, yet its simulation on quantum computers is fundamentally limited by the unitary nature of quantum evolution. 
We propose the quantum Koopman method, a data-driven framework that embeds nonlinear dynamics into a learned linear representation and implements the resulting evolution using shallow quantum circuits. This method learns Koopman observables from trajectory data, projects the lifted dynamics onto a finite-dimensional subspace, and decomposes the corresponding non-unitary propagator into parallel spectral channels. We utilize the Koopman method on a superconducting processor to simulate three distinct nonlinear systems, comprising reaction-diffusion dynamics, fluid motion on a sphere, and satellite-derived observations of Gulf Stream currents, employing up to 32 parallel circuits of 10 qubits. These quantum simulations capture the dominant multiscale patterns and statistical signatures of the underlying dynamics, and reveal a transition from performance limited by hardware noise in weakly nonlinear systems to performance limited by finite-dimensional Koopman representations as nonlinear scale interactions increase. This transition identifies a practical boundary for quantum-amenable nonlinear dynamics, establishing a hardware-validated route for simulating moderately nonlinear dynamics on near-term quantum hardware. 
\end{abstract}

\maketitle

\let\oldaddcontentsline\addcontentsline
\renewcommand{\addcontentsline}[3]{}

%\section*{Introduction}\label{sec:intro}

\noindent 
Quantum computing promises exponential speedups for the scientific simulation of complex systems~\cite{Feynman1982Simulating, Daley2022Practical}. 
However, quantum state evolution is governed by unitary operators, rendering these dynamics strictly linear and reversible~\cite{nielsen2010}. 
This intrinsic linearity stands in sharp contrast to the nonlinear dynamics ubiquitous in real-world phenomena, ranging from planetary-scale ocean circulation~\cite{beech2022long} to microscale kinetics in engine combustion~\cite{Givi2020Quantum}.  
Bridging this divide is essential for extending quantum utility to nonlinear regimes~\cite{Succi2023Quantum, Meng2024Challenges, Tennie2025Quantum}, yet any viable framework must reconcile mathematical rigor with the resource constraints of current quantum hardware~\cite{Hoefler2023Disentangling, aaronson2026future}.
Efficiently embedding nonlinear, dissipative dynamics within the unitary operations of quantum processors therefore remains a fundamental open challenge for the quantum simulation of real-world physical problems.

Existing quantum approaches to nonlinear dynamics generally fall into two broad families. 
The first follows a three-stage pipeline consisting of analytical linearization~\cite{liu2021efficient, joseph2020koopman, jin2023time, Succi2024Ensemble, alipanah2025quantum}, mapping onto a quantum linear-system or Hamiltonian-simulation algorithms~\cite{harrow2009quantum, childs2017quantum, An2023Linear, jin2024quantumPRL, Lu2024Quantum, Brearley2024Quantum}, and compilation into executable circuits~\cite{meng2025geometric}, with several alternative methods bypassing explicit linearization~\cite{meng2023quantum, meng2024quantum, Wang2025, lee2026multiple}. 
However, this family remains constrained by two limitations. 
First, the three stages are typically developed in isolation, yielding circuits whose depth exceeds what current noisy intermediate-scale quantum (NISQ)~\cite{Preskill2018Quantum} devices can reliably execute~\cite{aaronson2026future,mele2026noise}.
Second, the underlying analytical linearizations typically converge only under weak nonlinearity, leaving moderately nonlinear dynamics of real-world systems out of reach. 
The second family comprises variational quantum algorithms~\cite{biamonte2017quantum, lubasch2020variational, cerezo2022challenges, pfeffer2023reduced,jaksch2023variational}, which treat the quantum circuits as trainable, high-dimensional regressors. 
Without a structural connection to the underlying physics, these problem-agnostic circuit ansatzes lack the inductive bias needed to navigate complex parameter landscapes and frequently encounter barren plateaus~\cite{larocca2025barren} and local minima~\cite{anschuetz2022quantum}. 
Consequently, hardware demonstrations of these methods have been limited to 1D and 2D benchmarks using no more than 10 qubits~\cite{bharadwaj2023hybrid, wright2024noisy, Meng2024Simulatinga, chen2024enabling, wang2026simulating}. 
Although a recent Koopman-inspired approach~\cite{zhang2025data} has explored embedding physical priors into variational circuits, a unified framework that can simultaneously access moderately nonlinear regimes and compile into hardware-feasible circuits remains absent.

Here, we introduce the quantum Koopman method (QKM), which provides both a theoretical foundation and a hardware-efficient implementation for simulating nonlinear dynamics on an $n$-qubit quantum processor, as illustrated in Fig.~\ref{fig:schematic}.
Specifically, the Koopman representation lifts the nonlinear dynamics (Fig.~\ref{fig:schematic}a) into an infinite-dimensional linear space (Fig.~\ref{fig:schematic}c), which is subsequently projected onto a finite, $2^n$-dimensional subspace spanned by learned observables (Figs.~\ref{fig:schematic}b,d).
A suite of three theorems connects these non-unitary linearized dynamics to hardware-executable quantum circuits, yielding a topology-native ansatz derived from the linear combination of Hamiltonian simulation (LCHS)~\cite{An2023Linear} rather than heuristic design.
To prepare quantum states, a classical neural network (NN) encoder maps physical initial conditions to circuit parameters, achieving a $\mr{poly}(n)$-depth encoding by leveraging the regularity of physical fields (Fig.~\ref{fig:schematic}d).
The Koopman observables and operator parameters are learned jointly from data, enabling the direct execution of the resulting circuits on a superconducting quantum processor.

\begin{figure*} %[ht!]
    \centering
    \includegraphics{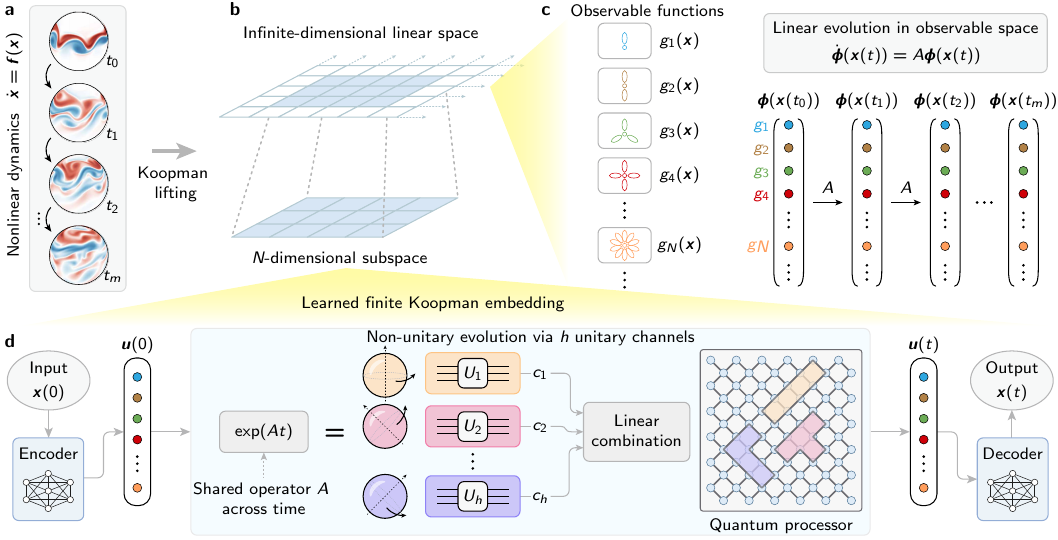}
    \caption{\textbf{Schematic of the QKM framework for simulating nonlinear dynamics on a superconducting quantum processor.}
    \textbf{a}, A nonlinear dynamical system $\dot{\boldsymbol{x}}=\boldsymbol{f}(\boldsymbol{x})$ generates discrete spatiotemporal snapshots $\boldsymbol{x}(t_0),\boldsymbol{x}(t_1),\cdots,\boldsymbol{x}(t_m)$.
    \textbf{b}, The nonlinear dynamics are mapped to an infinite-dimensional linear observable space via Koopman lifting in \textbf{c}, and are subsequently projected onto a finite $N$-dimensional subspace spanned by learned observables in \textbf{d}, yielding the linear evolution $\dot{\vec{u}}= A \vec{u}$.
    \textbf{c}, In full Koopman theory, this infinite-dimensional space comprises all observables $g_j(\boldsymbol{x})$ acting on the physical state, such that $\boldsymbol{\phi}(\boldsymbol{x})=(g_1(\boldsymbol{x}),g_2(\boldsymbol{x}),\cdots)\T$.
    Although the physical field $\boldsymbol{x}(t)$ evolves nonlinearly, the lifted observables evolve linearly under the Koopman generator according to the equation $\dot{\boldsymbol{\phi}}(\boldsymbol{x}(t))=A\boldsymbol{\phi}(\boldsymbol{x}(t))$.
    \textbf{d}, A finite Koopman embedding is learned from the data in \textbf{a}, yielding a low-dimensional latent state $\boldsymbol{u}(t)$ that preserves the dominant dynamics.
    These learned observables define a finite-dimensional representation in which the same linear generator $A$ propagates the state through time.
    In practice, the physical input $\boldsymbol{x}(0)$ is encoded into the finite Koopman state $\boldsymbol{u}(0)$, evolved under the non-unitary propagator $\mathrm{e}^{At}$, and decoded to reconstruct the physical state $\boldsymbol{x}(t)$.
    The non-unitary propagator $\mathrm{e}^{At}$ is implemented via the LCHS, which decomposes the propagator $\mathrm{e}^{At}$ into $h$ weighted unitary channels (exemplified by the yellow, red, and purple components).
    These $h$ independent circuits, characterized by distinct ring-topology qubit layouts, are executed in parallel on a superconducting quantum processor.
    }
    \label{fig:schematic}
\end{figure*}

We validate the QKM across three progressively challenging regimes.
These benchmarks comprise a reaction-diffusion system on a 3D cubic grid, shallow-water dynamics on a curvilinear sphere, and satellite-derived observations of the Gulf Stream. 
Our experiments utilize up to 32 parallel circuits of 10 physical qubits each, corresponding to $10^5$ computational grid points.
This effort represents a significant expansion in the scale of quantum resource integration for classical nonlinear dynamics, extending prior demonstrations that have been largely restricted to 1D and 2D benchmarks on fewer qubits. 
Together, these theoretical and experimental advances delineate the practical boundary of quantum utility for simulating nonlinear dynamics. 
The QKM operates effectively in the regime where the finite-dimensional Koopman representation and the shallow LCHS-derived ansatz remain faithful, attaining a speedup ratio of $\mc{S}\sim\mc{O}(2^n/n^3)$ over classical Koopman propagation. 
Our experiments demonstrate a transition from hardware-noise-limited to theory-limited accuracy, which operationally defines the boundary of this regime. 

\firtitle{Results}
\sndtitle{Quantum-circuit realization of Koopman dynamics}
The Koopman representation lifts the original nonlinear dynamics into an $N$-dimensional linear system $\dif  u\lrr{t}/\dif t =  A u\lrr{t}$, where $A\in\mbb{C}^{N \times N}$ is the finite-dimensional approximation of the Koopman operator and $ u$ denotes the vector of learned observables (see Methods). 
The resulting propagator $\ee^{A t}$ is generically non-unitary, as the eigenvalues of $A$ can exhibit non-zero real parts associated with dissipative or unstable dynamics.
To bridge the gap between this non-unitary evolution and hardware-executable quantum circuits, we establish a chain of three theorems that progressively transform  $\ee^{A t}$ into an ensemble of shallow quantum circuits (see Methods).
As illustrated in Fig.~\ref{fig:schematic2}, Theorems~\ref{theory:1} (diagonalized LCHS), \ref{theory:2} (spectral sampling convergence), and \ref{theory:3} (universal approximation bound for diagonal unitaries) establish a framework for the efficient simulation of nonlinear dynamics via $h$ independent quantum circuits, where each circuit requires only a single layer of $R_z$ for time evolution. 
With the $N$-dimensional Koopman state encoded into $n=\log_2 N$ qubits, explicit classical propagation of $\ee^{At}u\lrr{0}$ requires at least $\mc{O}(N)=\mc{O}(2^n)$ operations.
The QKM replaces this dense-vector propagation by $h$ shallow circuits with $\mc{O}(n)$ gates per channel, yielding an evolution-step cost of $\mc{O}(hn)$ and a corresponding speedup over classical Koopman propagation.

\begin{figure*} %[ht!]
    \centering
    \includegraphics{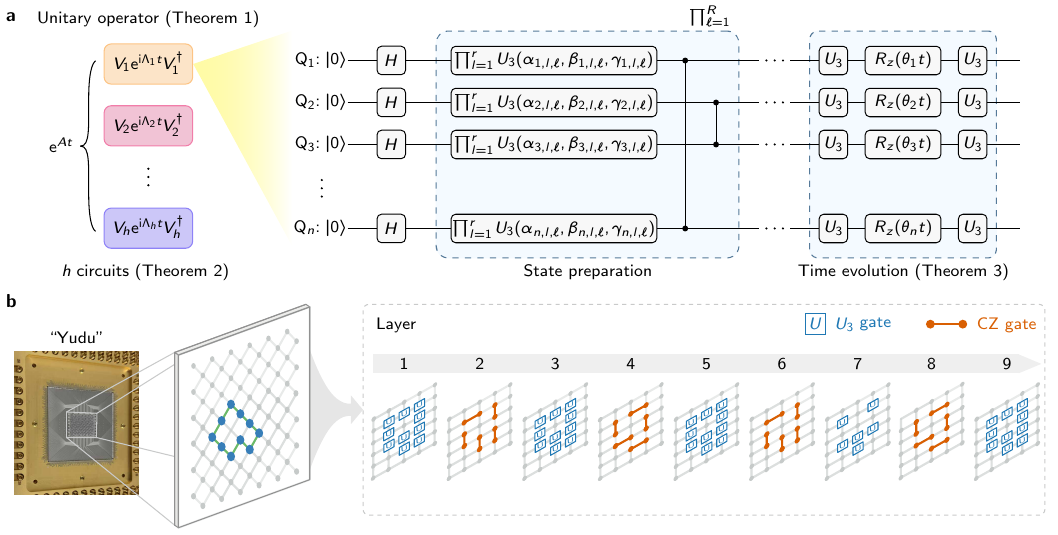}
    \caption{\textbf{Hardware-efficient quantum circuit of the QKM.} 
    \textbf{a}, The non-unitary propagator $\ee^{At}$ for nonlinear dynamics is first reformulated as a spectral integral of diagonal unitary operators (Theorem~\ref{theory:1}).
    Each of the $h$ independent PQCs implements a single spectral component $V(k)\ee^{\ii\Lambda(k)t} V^\dagger(k)$ of the diagonalized LCHS decomposition (Theorem~\ref{theory:2}). 
    Each $n$-qubit circuit comprises two functional blocks.
    The state-preparation block utilizes $R$ alternating layers of parameterized $U_3$ gates with rotation angles $\{ \alpha_{j,l,\ell}, \beta_{j,l,\ell}, \gamma_{j,l,\ell} \}$ generated by the classical encoder, and $\mathrm{CZ}$ pairs that alternate between odd layers $\{(n,1),(2,3), \dots, (4,5)\}$ and even layers $\{(1,2),(3,4), \dots, (n-1,n)\}$, forming a parity-based ring topology.
    The time-evolution block implements a sandwich ansatz that nests a central $R_z$ layer (Theorem~\ref{theory:3}), which encodes the diagonal eigenvalues $ \Lambda(k)$, between $U_3$ blocks realizing the basis transformation $ V(k)$ and its adjoint $ V^\dagger(k)$.
    \textbf{b}, Hardware-native compilation of the QKM circuit on the superconducting processor ``Yudu''.
    The circuit ansatz in \textbf{a} is mapped onto a connected 10-qubit subgraph of the device such that all two-qubit gates respect the native nearest-neighbour couplings of the chip.
    Starting from the ground state $\ket{0}^{\otimes 10}$, we apply parallel single- and two-qubit gates layer by layer up to a circuit depth of nine.
    In practice, each layer of parallel single-qubit (two-qubit) gates requires 40 (90) ns.
    Consequently, the total execution time is 560 ns, which is substantially shorter than the median qubit lifetime 48$\mu$s. 
    }
    \label{fig:schematic2}
\end{figure*} 

To complete the simulation pipeline, the quantum evolution must be preceded by state preparation. 
Preparing an arbitrary $2^n$-dimensional quantum state requires $\mc{O}(2^n)$ quantum resources ~\cite{zhang2022quantum, sun2023asymptotically}, which would eliminate the advantage in the evolution step.
Fortunately, most physically relevant initial conditions are not arbitrary and possess additional spatial structure, such as smoothness and symmetry~\cite{meng2025geometric}. 
We exploit this structure by employing a classical NN encoder that maps physical states to the rotation angles of a parameterized quantum circuit (PQC) of depth $\mc{O}(\mr{poly}(n))$.
The universal approximation capability of NNs~\cite{hornik1989multilayer} ensures that such a mapping is learnable, while the universality of the gate set~\cite{barenco1995elementary} guarantees sufficient expressivity. 
A complementary oracle-based analysis in Sec.~\ref{sec:prep} in Supplementary Information (SI)~\cite{SI} justifies the efficient state-preparation assumption for structured physical initial conditions. Numerical verification of this expressibility is provided in Sec.~\ref{sec:err_prep} in SI~\cite{SI}.

\vspace{1em}
\sndtitle{Hardware-efficient end-to-end implementation}
The QKM framework is realized through a hybrid procedure comprising a classical encoder-decoder pair and $h$ parallel PQCs in Fig.~\ref{fig:schematic}d.
Given an initial condition $ x(0)$, the encoder maps it to $h$ sets of rotation angles, with each set preparing an $n$-qubit quantum state on its respective PQC.
Each PQC then applies a time-evolution block to implement a single spectral component $ V(k)\ee^{\ii\Lambda(k)t} V^\dagger(k)$ of the diagonalized LCHS in Eq.~\eqref{eq:LCHS}.
Following computational-basis measurements, the shot-count distributions from the $h$ circuits are processed by the decoder to reconstruct the evolved state $ x(t)$. 
This workflow directly instantiates our theoretical framework: the encoder leverages physical input regularity to ensure efficient state preparation; the time-evolution block realizes the unitary representation of $\ee^{At}$ guaranteed by Theorems~\ref{theory:1} and \ref{theory:2}; and the single-layer $R_z$ configuration yields the shallow circuit prescribed by Theorem~\ref{theory:3}.
Comprehensive details regarding the encoder, decoder, and circuit construction are provided in Methods and Secs.~\ref{sec:algorithm} and \ref{sec:AE} in SI~\cite{SI}.

\begin{table*}[ht!]
	\centering
    \setlength{\tabcolsep}{8pt}
    \renewcommand{\arraystretch}{1.2}
	\caption{\textbf{Quantum simulation parameters.}  
    Summary of simulation settings across the three cases, detailing the classical grid resolution $\mc{D}$ and the structural parameters of the $h$ PQCs, including the qubit count $n$, circuit depth $R$, and $U_3$ gate density $r$. 
    The theoretical evolution speedup scales as $\mc{S}_\text{evo} = 2^n/(hn)$, while the number of measurement repetitions is determined by the shot count $M$.
    Experimental metrics for the total circuit depth $L$ and gate count $G$ are utilized to evaluate the quantum speedup $\mc{S} = 2^n / (hG)$.
    In experiments, the relative theoretical error is quantified by the training loss $\varepsilon_\text{th}\approx\ell_\text{train}$.
    }
	\label{tab:setup}
	\begin{tabular}{lccccccccccc}
    \hline\hline 
	Case & $\mc{D}$  & $n$ & $R$ & $r$ & $h$ & $M$ & $L$ & $G$ & $\varepsilon_\text{th}$ & $\mc{S}_\text{evo}$ & $\mc{S}$ \\
    \hline
    3D reaction-diffusion & $64\times64\times64$ & 6 & 4 & 1 & 8 & 6144 & 12 & 58 & 0.001 & 1.33 & 0.14 \\
	Spherical fluid & $512\times256$ & 10 & 3 & 3 & 32 & 10240 & 17 & 146 & 0.015 & 3.2 & 0.22 \\
    Ocean currents & $256\times256$ & 10 & 3 & 3 & 8 & 10240 & 17 & 145 & 0.020 & 12.8 & 0.88 \\
    \hline\hline
	\end{tabular}
\end{table*}

Crucially, the circuit structure is dictated by the LCHS decomposition rather than an empirical hardware-efficient ansatz.
This systematic construction provides interpretability and error traceability while facilitating the training of the QKM. 
The state-preparation block (Fig.~\ref{fig:schematic2}a) comprises $R$ alternating layers, each consisting of $r$ single-qubit gates followed by $\mathrm{CZ}$ entangling gates.
Each time-evolution block (Fig.~\ref{fig:schematic2}a) features a sandwich architecture, wherein a central layer of $R_z$ rotations encoding the diagonal $\Lambda(k)$ is flanked by two $U_3$ blocks realizing the basis transformation $ V(k)$ and its adjoint.
In the transpiled circuit, the $\mathrm{CZ}$ gates alternate between odd and even layers across adjacent qubits, forming a ring topology that maps directly onto the connectivity of the superconducting processor ``Yudu'' in Fig.~\ref{fig:schematic2}b.
This hardware-native entangling layout eliminates SWAP insertions during transpilation and compresses the post-transpilation circuit depth by up to 50\% (see Methods), which is critical for preserving simulation fidelity on NISQ devices.
Furthermore, the parallel-circuit structure guaranteed by Theorem~\ref{theory:2} can be preserved at the hardware level, with the $h$ PQCs executed concurrently on disjoint qubit subsets of the processor.

To accommodate NISQ-era constraints, we set $n=\mc{O}(\log \mc{D})$, $h=\mc{O}(n)$, and $Rr=\mc{O}(n)$ for a nonlinear system with $\mc{D}$ grid points, yielding $\mc{O}(n)$-depth state preparation.
The choice of $h$ balances the spectral approximation bound of Theorem~\ref{theory:2} against resource overhead, whereas $Rr$ balances state-preparation expressibility against gate-error accumulation.
Both trade-offs are detailed in Secs.~\ref{sec:ablation_h} and~\ref{sec:err_prep} in SI~\cite{SI}.
The specific values of $n$, $h$, $R$, and $r$ employed in each benchmark are summarized in Tab.~\ref{tab:setup}. 

\vspace{1em}
\sndtitle{Simulation case setup and hardware overview} 
We evaluate the QKM across three physically distinct and progressively challenging regimes, each executed on the superconducting processor ``Yudu''~\cite{BAQIS2024Quafu}.
The first is a 3D reaction-diffusion system governed by the Gray-Scott equations on a cubic domain, where the competition between diffusion and nonlinear reaction yields complex, self-organizing spatiotemporal patterns.
The second involves the shallow-water equations on a sphere, a canonical model in geophysical fluid dynamics, where multiscale rotational dynamics impose stringent demands on simulation accuracy.
The third moves beyond prescribed equations and applies the QKM to satellite-derived Gulf Stream observations, converting real geophysical data into a quantum-executable model of ocean-current dynamics.

The superconducting processor ``Yudu'' comprises 72 frequency-tunable transmon qubits arranged in a lattice geometry, with comprehensive device specifications provided in Sec.~\ref{sec:device} in SI~\cite{SI}.
In this work, we utilize up to 10 qubits per subcircuit and up to 32 parallel subcircuits to perform the experiments.
The target experimental circuits are implemented using the native gate set $\{U_3, \mathrm{CZ}\}$.
Through optimized control procedures, we achieve parallel single- and two-qubit gate fidelities of 99.92\% and 98.82\%, respectively. 
Experimental results are reconstructed by the NN decoder from the computational-basis measurements on all qubits of the $h$ parallel circuits, with each circuit executed for $M$ shots. 
The reference solutions are obtained from direct numerical simulations for the 3D reaction--diffusion and spherical fluid dynamics cases, and from satellite-derived altimetry observations for the Gulf Stream case. 
Tab.~\ref{tab:setup} summarizes the problem sizes, circuit configurations and experimental parameters used in the three benchmarks.

\vspace{1em}
\sndtitle{3D reaction-diffusion systems}
The first test case targets the 3D Gray-Scott reaction-diffusion system~\cite{gray1983autocatalytic} (see Sec.~\ref{sec:cube} in SI~\cite{SI}). 
In this system, the competition between chemical reaction and spatial diffusion triggers the emergence of complex, self-organizing dissipative structures.
This benchmark assesses the capacity of the QKM to encode high-dimensional ($\mc{D}=64^3$) nonlinearities into a compact quantum representation using $n=6$ qubits per circuit.

\begin{figure*}
    \centering
    \includegraphics{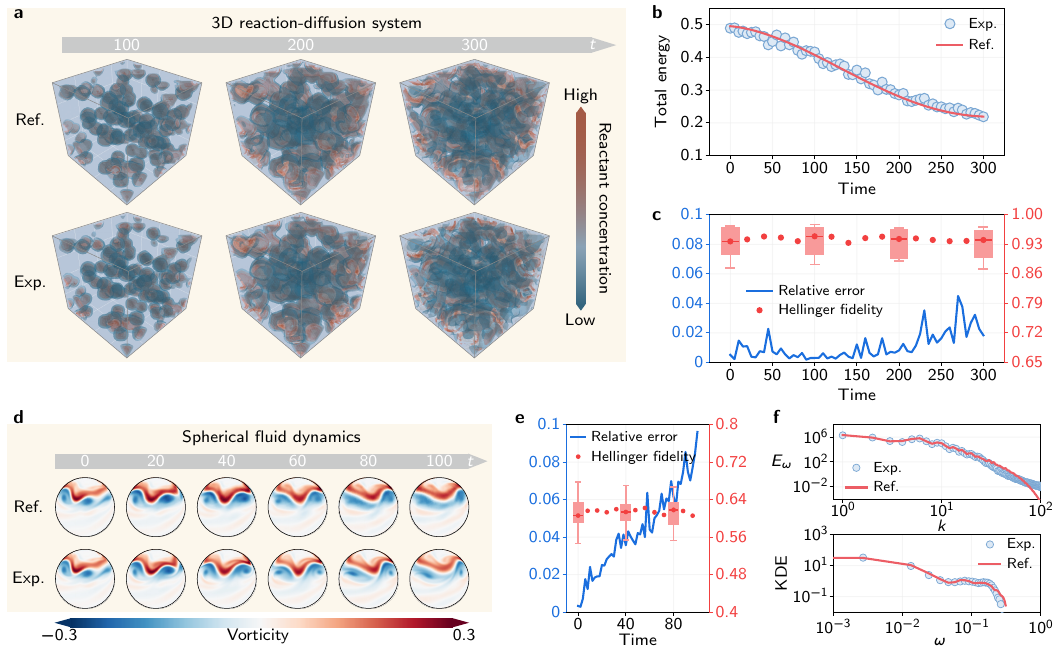}
    \caption{\textbf{Assessment of the QKM framework on benchmark nonlinear dynamical systems.}
    \textbf{a}, Quantum simulation of a 3D reaction-diffusion system using the QKM on the superconducting processor ``Yudu''.
    Volume renderings compare the experimental reactant-concentration fields $u$ with the reference solutions at $t=100$, 200, and 300.
    The QKM result accurately captures the temporal evolution of the total energy $\langle u^2 \rangle / 2$ in \textbf{b}, where $\langle\cdot\rangle$ denotes volume average. 
    The relative error $\varepsilon_{L2}$ remains small, and the median Hellinger fidelity stays close to unity in \textbf{c}, demonstrating quantitative agreement with the nonlinear dynamics.
    \textbf{d}, Quantum simulation of spherical fluid dynamics using the QKM on the superconducting processor ``Yudu''.
    The experimental vorticity fields $\omega$ are compared with the reference solutions at $t=0$, 20, 40, 60, 80, and 100.
    The relative error remains below 10\%, and the median Hellinger fidelity stays near 0.65 throughout the simulation in \textbf{e}.
    The comparison of the experimental enstrophy spectrum $E_\omega$ and vorticity KDE with the reference solutions at $t=80$ in \textbf{f} confirms the accuracy of the QKM.
    }
    \label{fig:cubesphere}
\end{figure*}

The QKM successfully simulates the 3D reaction-diffusion dynamics, characterized by the emergence, growth, and coalescence of self-organized dissipative structures of the reactant-concentration field $u$, as shown in Fig.~\ref{fig:cubesphere}a. 
Quantitatively, the total energy tracks the dissipative relaxation of the system (Fig.~\ref{fig:cubesphere}b), and the relative error $\varepsilon_{L2}$ remains below 0.05 and the median Hellinger fidelity is maintained near 0.93 throughout the evolution (Fig.~\ref{fig:cubesphere}c).
These results demonstrate that the diagonalized unitary dynamics established in Theorems~\ref{theory:1} to \ref{theory:3} enable physically consistent 3D spatiotemporal simulations on a NISQ device.

\vspace{1em}
\sndtitle{Spherical fluid dynamics}
The second benchmark extends the evaluation to spherical fluid dynamics governed by the shallow-water equations~\cite{galewsky2004initial} (see Sec.~\ref{sec:sphere} in SI~\cite{SI}).
This case introduces topological constraints absent in Euclidean domains, specifically periodic longitudinal closure and polar coordinate singularities.
The resulting dynamics, which include mid-latitude jet instability, planetary-scale vortex shedding, and turbulent spectral cascades, exhibit a heightened sensitivity to phase errors and discretization artifacts compared to their flat-domain counterparts.
At a grid resolution of $\mc{D}=512\times 256$ encoded using $n=10$ qubits per circuit and $h=32$ parallel PQCs, this case constitutes the largest circuit deployment and the most geometrically complex test within this study.

The QKM simulates the long-time vorticity evolution under spherical shallow-water dynamics in Fig.~\ref{fig:cubesphere}d.
The quantum simulation captures the growth of the mid-latitude instability and the emergence of large-scale coherent vortices on the sphere, while quantum noise manifests primarily as localized fluctuations in the vorticity snapshots.
As illustrated in Fig.~\ref{fig:cubesphere}e, the median Hellinger fidelity stabilizes near $0.6$, whereas the relative error grows at a moderate rate. 
The enstrophy spectrum $E_\omega$ closely matches the multiscale distribution of vorticity across wavenumbers, while the kernel density estimate (KDE) of the vorticity recovers consistent one-point statistics, as shown in Fig.~\ref{fig:cubesphere}f.
These diagnostic assessments demonstrate that the QKM maintains the core vortical, spectral, and statistical properties of spherical fluid dynamics. 

Moreover, the close agreement of the enstrophy spectrum over a broad range of scales indicates that the quantum simulation preferentially preserves the core dynamics of the system.
The principal low-wavenumber modes containing the bulk of the enstrophy are encoded with high fidelity, whereas discrepancies are restricted to the high-wavenumber tail where both Koopman truncation and quantum noise accumulate.
This spectral selectivity is consistent with the complexity analysis, which assumes that many physically relevant systems are dominated by low-frequency components that admit efficient Koopman truncation.
A comparison with a noiseless simulation indicates that the theoretical error $\varepsilon_{\mr{th}}$ and hardware noise $\varepsilon_{\mr{noise}}$ are of comparable magnitude in this case (detailed in Sec.~\ref{sec:sim_sphere} in SI~\cite{SI}).

\vspace{1em}
\sndtitle{Real-world ocean currents}
The final benchmark evaluates the QKM on the complex mesoscale dynamics of the Gulf Stream using real-world observational data~\cite{CMEMS} (see Sec.~\ref{sec:square} in SI~\cite{SI}). 
Unlike previous benchmarks governed by closed-form partial differential equations under controlled initial conditions, this scenario utilizes satellite-derived altimetry products characterized by observational gaps and irregular land-sea boundaries. 
Simulating this system tests the capacity of the QKM to function not merely as an equation solver, but as a data-driven framework that assimilates observational data into a quantum-executable form. 
The Gulf Stream region ($20^{\circ}$N to $52^{\circ}$N, $33^{\circ}$W to $65^{\circ}$W) is discretized at a resolution of $\mc{D}=256\times 256$ and encoded using $n=10$ qubits with $h=8$ parallel PQCs. 

\begin{figure}
    \centering
    \includegraphics[scale=1.0]{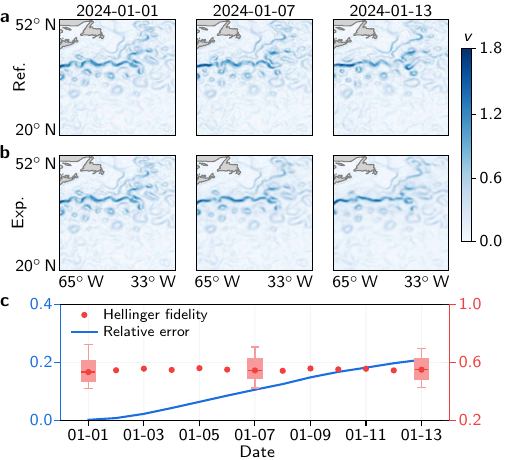}
    \caption{\textbf{Quantum simulation of real-world Gulf Stream ocean currents on the superconducting processor ``Yudu''.}
    \textbf{a}, Surface geostrophic velocity $v$ derived from satellite observations on three dates spanning January 1 to 13, 2024.
    \textbf{b}, Corresponding QKM results simulated by the ``Yudu'' processor.
    \textbf{c}, Temporal evolution of the relative error $\varepsilon_{L2}$ and the median Hellinger fidelity, with box plots showing the fidelity distribution across circuit instances on January 1, 7, and 13, 2024.
    }
    \label{fig:square}
\end{figure}

A comparison between the satellite-derived data in Fig.~\ref{fig:square}a and the QKM results in Fig.~\ref{fig:square}b demonstrates that the QKM successfully simulates the principal mesoscale features of the surface geostrophic velocity field $v$, including jet and eddy structures, over a 13-day evaluation period in January 2024.  
As shown in Fig.~\ref{fig:square}c, the error $\varepsilon_{L2}$ exhibits a gradual growth characteristic of chaotic dynamical systems, yet remains bounded at approximately $0.2$ after nearly two weeks of evolution.
Throughout the evaluation window, the median Hellinger fidelity $F$ remains stable at approximately $0.6$, indicating robust statistical agreement between the hardware-sampled and noiseless quantum distributions. 

These findings demonstrate that the QKM possesses the expressive capacity and noise resilience required to manage the inherent uncertainties of real-world geophysical observations.  
The growth of the $L_2$ error, in contrast with the temporally stable Hellinger fidelity, suggests that while the quantum processor continues to faithfully execute the learned circuits, the finite-dimensional Koopman observable space gradually loses its capacity to track the chaotic trajectory over extended horizons.  
Consequently, expanding the dimension $N$ of the observable space represents the primary avenue for improving long-term predictive accuracy. 

Together, these three benchmarks demonstrate that the QKM provides a unified and hardware-validated framework for the digital quantum simulation of nonlinear dynamics. 
The QKM maintains physically consistent accuracy on a superconducting processor without requiring modification to the circuit topology or the LCHS formulation, confirming the portability and practical viability of the underlying framework. 

\vspace{1em}
\sndtitle{Pathway to quantum utility with QKM}
Although the QKM speedup $\mc{S}\sim \mc{O}(2^n/n^3)$ derived in Methods (blue line in Fig.~\ref{fig:path}) is attainable for any nonlinear system using the single-layer $R_z$ ansatz, the accuracy at which this ceiling is reached varies across systems.
The performance metric is the relative theoretical error $\varepsilon_{\text{th}}$, which combines the approximation error of the finite-dimensional Koopman representation and that of the single-layer $R_z$ ansatz. 
In our experiments, the ansatz approximation error $\varepsilon_{\text{ansatz}}$ due to the limited dimension $N$ dominates $\varepsilon_{\text{th}}$ (see Sec.~\ref{sec:ablation_rzz} in SI~\cite{SI}). 
Theorem~\ref{theory:3} therefore establishes the scalability limit, bounding $\varepsilon_{\text{ansatz}}$ by the high-order Pauli-$Z$ string coefficients of the target Hamiltonian. 
Retaining higher-order terms enriches ansatz and reduces $\varepsilon_{\mr{th}}$ by capturing additional spectral content, albeit at the expense of a reduced $\mc{S}$. 
The choice of ansatz therefore defines an accuracy-speedup trade-off curve, ranging from the shallow limit at $\mc{S}\sim \mc{O}(2^n/n^3)$ to the deep limit of exact representation at $\mc{S}\sim \mc{O}(1/n)$. 
While any system can be simulated using shallow circuits, complex systems must trade speedup $\mc{S}$ for accuracy $\varepsilon_\text{th}$.
We adopt a threshold of $\varepsilon^* = 10^{-3} $ on $\varepsilon_{\text{th}}$ as the operational criterion, quantified in practice by the training loss $\ell_{\text{train}}$.

\begin{figure}
    \centering
    \includegraphics{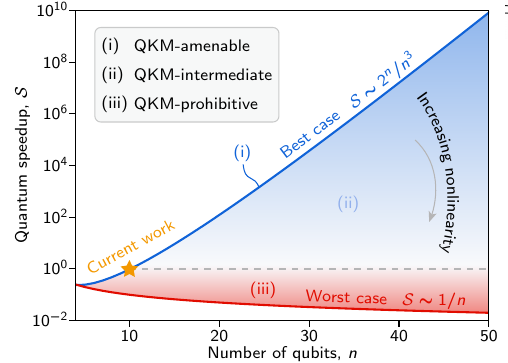}
    \caption{\textbf{Quantum-speedup regimes of the QKM.} 
    The quantum speedup $\mathcal{S}$ as a function of the number of qubits $n$. 
    The upper blue curve represents the best-case scaling, $\mathcal{S}\sim 2^n/n^3$, achieved when a first-order circuit ansatz corresponding to a single $R_z$ layer already provides sufficient accuracy. 
    Problems in this limit are classified as QKM-amenable. 
    With increasing nonlinearity of the target dynamics, higher-order QKM representations and more retained terms are required, thereby increasing the quantum cost and lowering the achievable speedup.
    The region between the best-case curve and the threshold $\mathcal{S}=1$ defines the QKM-intermediate regime, where the first-order approximation is insufficient, yet the quantum implementation remains faster than the corresponding classical Koopman propagation.
    The dashed line $\mathcal{S}=1$ marks the crossover between quantum advantage and the absence of advantage.
    Below this threshold, problems enter the QKM-prohibitive regime, where the required accuracy approaches the full expansion limit, the worst-case scaling (red curve) decreases to $\mathcal{S}\sim 1/n$. 
    The star indicates the scale of the present experiments with $n=10$, near the $\mathcal{S}=1$ utility threshold.
    }
    \label{fig:path}
\end{figure}

This trade-off partitions nonlinear dynamical systems into three distinct regimes based on their spectral compressibility under a learned Koopman embedding. 
For QKM-amenable systems, the single-layer $R_z$ ansatz achieves an error of $\varepsilon_{\text{th}} \lesssim \varepsilon^*$, enabling the speedup $\mc{S}\sim \mc{O}(2^n/n^3)$ to be realized at scientific-computing accuracy. 
For QKM-intermediate systems, the shallow ansatz yields $\varepsilon_{\text{th}} > \varepsilon^*$, but augmenting the ansatz with $R_{zz}$ and higher-order gates reduces $\varepsilon_{\text{th}}$ at a polynomial cost in $n$. 
In this regime, quantum advantage remains theoretically viable at a reduced speedup and becomes practically accessible once gate fidelities improve sufficiently to support the deeper circuits.  
Conversely, QKM-prohibitive systems are characterized by intrinsically incompressible spectral dynamics, so that no finite-dimensional Koopman embedding can concentrate their spectral weight into low-order modes. 
A canonical example is fully-developed turbulence~\cite{zhang2025data}, in which the energy cascade distributes spectral weight across all wavenumbers. 
Capturing such dynamics demands an ansatz whose cost scales exponentially with $n$, degrading the worst-case performance to $\mc{S}\sim\mc{O}(1/n)$ (red line in Fig.~\ref{fig:path}) and eliminating the quantum advantage. 
Representing these systems instead requires tailored methods focused on extracting dominant coherent structures~\cite{meng2025geometric}.

The three benchmarks calibrate this partition. 
With measured training losses $\ell_{\text{train}}$ ranging from $10^{-3}$ to $2\times 10^{-2}$ (see Tab.~\ref{tab:setup}), the 3D reaction-diffusion benchmark falls within the QKM-amenable regime, whereas the spherical shallow-water and Gulf Stream benchmarks lie in the QKM-intermediate regime. 
Comparing hardware execution against a noiseless simulation (Sec.~\ref{sec:sim} in SI~\cite{SI}) resolves the dominant error source in each case, revealing a progression from hardware-noise-limited to theory-limited performance as nonlinearity grows. 
Critically, by learning observables directly from data rather than relying on an analytical expansion, the QKM maps dynamics with broadband spectral content into the reach of shallow circuits, including mesoscale geophysical flows that are beyond the scope of truncation-based linearization. 
The framework thereby establishes a broader operational envelope for simulating general nonlinear problems. 

\firtitle{Discussion}
The QKM unifies the Koopman operator theory, Hamiltonian simulation, and hardware-native compilation into a single jointly trained pipeline, recasting non-unitary Koopman dynamics into shallow parallel circuits on a superconducting processor. 
The scaling of speedup ratio $\mc{S}\sim\mc{O}(2^n/n^3)$ establishes the performance ceiling of the QKM; the regime criterion $\varepsilon_{\mr{th}}\lesssim\varepsilon^*$ specifies when a given system achieves this bound under the single-layer $R_z$ ansatz. 
Implemented on up to 32 parallel 10-qubit circuits, the 3D reaction-diffusion benchmark falls within the QKM-amenable regime, establishing that this performance ceiling is not merely asymptotic but practically reachable on NISQ devices. 
The spherical shallow-water and Gulf Stream benchmarks exceed this threshold under the shallow ansatz, thereby mapping the boundary behavior predicted by the QKM and indicating where richer ansatzes are required to further reduce $\varepsilon_{\mr{th}}$. 

These results are achieved using a structured circuit ansatz derived from the LCHS decomposition rather than generic variational heuristics.
The global representational capacity of the QKM stems from integrating Koopman lifting and an NN encoder, which jointly learn a finite-dimensional observable subspace that linearizes the dynamics.
The LCHS decomposition then translates this representation into physically motivated circuits, while a topology-native layout preserves this structure on hardware without requiring SWAP insertions.
Together, these features provide interpretability and error traceability while encoding underlying physical structures beyond those of empirical ansatz.
Analytical embeddings, such as Carleman~\cite{liu2021efficient} and Koopman-von Neumann~\cite{joseph2020koopman,jin2023time} formulations, are typically restricted to regimes of weak nonlinearity and strong dissipation.
By contrast, learning observables from data identifies an invariant subspace adapted to the dynamics, extending the accessible range to the moderately nonlinear regimes demonstrated in this work.

However, several limitations remain. 
First, the quantum-classical input/output bandwidth is the primary obstacle to scaling up the practical speedup. 
The input/output latency and noise accumulation preclude efficient on-device training, so the QKM is trained classically in a one-time offline preprocessing phase before deployment on quantum hardware. 
This restriction constrains the accessible observable dimension, limiting the present implementation to reduced-dimensional demonstrations and deferring the validation of practical advantages at larger qubit scales to future studies.
Second, strongly multi-scale dynamics, such as turbulence~\cite{gourianov2022quantum,li2024learning}, challenge both the shallow ansatz and the encoder, because high-order Pauli-$Z$ interactions increase the ansatz error while the smooth bias of the NN encoder can attenuate fine-scale information.
Finally, establishing rigorous bounds on the projection error remains an open challenge for data-driven Koopman methods.

Extending the operational reach of the QKM requires the coordinated development of both hardware and algorithms.
Specifically, improved gate fidelities enable the reliable execution of deeper circuits, while broader qubit connectivity facilitates native, multi-qubit entangling operations.
These hardware advancements pave the way for augmenting the time-evolution block with higher-order entangling gates, rendering the QKM-intermediate regime practically accessible. 
On the algorithmic side, learning observables that concentrate spectral weight into lower-order Pauli-$Z$ interactions can transition a QKM-intermediate system into the QKM-amenable regime, providing quantum utility without requiring deeper circuits. 
Finally, refining the theoretical bounds and exploring early fault-tolerant architectures will further clarify the boundary between QKM-amenable and QKM-prohibitive systems, ultimately enabling quantum utility in simulating complex, real-world physical phenomena.

\firtitle{Methods}
\sndtitle{Koopman operator theory}
The Koopman operator theory reformulates nonlinear dynamics as linear time evolution in an infinite-dimensional function space~\cite{Brunton2022Modern}.
Consider a smooth dynamical system $\dif x\lrr{t}/\dif t = f[ x\lrr{t}]$, where $ x \in \mc{X}$ represents the state on a manifold $\mc{X} \subset \mbb{R}^\mc{D}$ embedded in the $\mc{D}$-dimensional real space $\mbb{R}^\mc{D}$, and $f$ is a nonlinear vector field.  
Its solution defines a flow map $\Phi^t \colon \mc{X} \to \mc{X}$, such that $ x(t) = \Phi^t[ x(0)]$ for an initial condition $ x(0)$. 

Instead of directly evolving the state $ x$, the Koopman framework considers the evolution of an observable $g\colon \mc{X} \to \mbb{C}$ belonging to a function space $\mc{G}(\mc{X})$, where $\mbb{C}$ denotes the complex space.
The Koopman operator $\mc{K}^t$ acts on this observable via $\mc{K}^t g\lrr{ x} = g[\Phi^t\lrr{ x}]$.
Although the flow map $\Phi^t$ may be nonlinear, $\mc{K}^t$ is linear by construction.
Consequently, the infinitesimal generator $\mc{A}$ of $\mc{K}^t$ induces a linear dynamical system $\dif g/\dif t=\mc{A}g$~\cite{wang2026quantum, Jennings2026}, where $\mc{A}$ represents the Lie derivative of $g$ along $f\lrr{x}$, defined as $\mc{A}g \coloneqq \lim_{t \to 0} (\mc{K}^t g - g)/t = \bn g \cdot f$.

Although the observable space $\mc{G}(\mc{X})$ is inherently infinite-dimensional, practical computation necessitates projecting the dynamics onto a finite-dimensional subspace.
Identifying the optimal dimensionality for such a projection remains an open challenge in Koopman analysis~\cite{lin2021data}.
In the QKM, the state $ x$ is mapped to a vector of $N$ basis functions $ u = [g_1\lrr{ x}, \cdots, g_N\lrr{ x}]\T \in \mbb{C}^N$ (see Fig.~\ref{fig:schematic}c), yielding an approximate linear dynamical system $\dif u\lrr{t}/\dif t =  A u\lrr{t}, u\lrr{0} =  u_0$, where $ A \in \mbb{C}^{N \times N}$ represents the finite-dimensional approximation of $\mc{A}$.
Both the observable functions $\{g_i\}$ and the operator $ A$ are learned jointly from data via end-to-end training of the NN encoder and the quantum circuit parameters.

\vspace{1em}
\sndtitle{Theorems for unitary mapping}\vspace{-1em}
\begin{theorem}[Diagonalized LCHS]\label{theory:1}
Let $ A \in \mbb{C}^{N \times N}$ be decomposed into Hermitian and anti-Hermitian parts, $ A= L+\ii H$, with $ L=( A+ A^\dagger)/2$ and $ H=( A- A^\dagger)/2\ii$. If $L\preceq 0$, then for $t \ge 0$, the propagator $\ee^{ A t}$ admits the spectral integral representation
\EQ\label{eq:LCHS}
    \ee^{A t} = \int_{\mbb{R}} \frac{1}{\pi(1+k^2)}  V(k) \ee^{\ii\Lambda(k)t} V^\dagger(k) \dif k, 
\EN
where $ H + k L = V(k)\Lambda(k) V^\dagger(k)$ is the spectral decomposition of the Hermitian operator $ H + k L$ for each $k \in \mbb{R}$. 
\end{theorem}

Theorem~\ref{theory:1} reformulates the non-unitary propagator as a spectral integral of diagonal unitary operators, demonstrating the feasibility of a unitary representation.
The diagonal matrix $\Lambda(k)$ reduces the time evolution within each spectral component to phase rotations on individual basis states, a simplification exploited in Theorem~\ref{theory:3} to guarantee a shallow circuit depth.
The condition $L \preceq 0$ is satisfied without loss of generality via a rescaling of variables (see Sec.~\ref{sec:theorem1} in SI~\cite{SI}).

\begin{theorem}[Spectral sampling convergence]\label{theory:2}
Let $p(k) = [\pi(1+k^2)]^{-1}$ be the Cauchy-Lorentz density and $U_k(t) = \ee^{\ii(H+kL)t} = V(k) \ee^{\ii\Lambda(k)t}V^\dagger(k)$ be the unitary operator defined in Theorem~\ref{theory:1}. 
For an $h$-point uniform approximation over the truncated interval $[-K, K]$ with spacing $\Delta k = 2K/h$, the error is bounded by
\begin{equation}\label{eq:th2}
    \| \ee^{At} - \sum_{j=1}^{h} \Delta k \cdot p(k_j) U_{k_j}(t) \|_2 \le \frac{3}{\pi}\bigg(\frac{3\sqrt{3}+8t \| L \|_2}{4h}\bigg)^{1/3},
\end{equation}
where the optimal truncation radius $K=\big(\frac{3\sqrt{3}+8t \| L \|_2}{4h}\big)^{-1/3}$ minimizes this upper bound.
\end{theorem}

Theorem~\ref{theory:2} replaces the spectral integral in Eq.~\eqref{eq:LCHS} with a finite sum over $h$ points (see Sec.~\ref{sec:theorem2} in SI~\cite{SI}), where $h$ serves as a hyperparameter controlling the approximation fidelity. 
This decomposition of a deep circuit into $h$ shallow ones is essential for near-term hardware.
Although the $h$ spectral components could be encoded into a single $\log_2(hN)$-qubit circuit through entanglement, doing so would demand significantly deeper circuits with controlled multi-qubit operations. 
Distributing the computation across $h$ independent $n=\log_2 N$ qubits keeps each circuit shallow. 
Since these circuits share no quantum state, they can be executed concurrently to disjoint regions of the processors, as illustrated in Fig.~\ref{fig:schematic}e.

\begin{theorem}[Universal approximation bound for diagonal unitaries]\label{theory:3}
Let $H^* \in \mathbb{C}^{2^n \times 2^n}$ be a diagonal Hamiltonian with the induced unitary $U^* = \ee^{\ii H^*}$. The single-layer $R_z$ subspace is defined as $\mc{F} = \{ \ee^{\ii\phi} \bigotimes_{j=1}^n R_z(\theta_j) \mid \theta_j, \phi \in \mathbb{R} \}$,
where $R_z(\theta) = \ee^{-\ii \sigma_z \theta /2}$ and $\sigma_z$ is the Pauli-$Z$ operator. For any $U \in \mc{F}$, the minimum approximation error satisfies
\EQ\label{eq:th3}
\min_{U \in \mc{F}} \frac{1}{2^n} \| U - U^* \|_F^2 = \sum_{|s| \ge 2} \alpha_s^2 + \mathcal{O}( \max_{|s| \ge 2} \alpha_s^4 )
\EN
where $s = s_1 \dots s_n \in \{0,1\}^n$ is a bitstring with Hamming weight $|s|$, $\|\cdot\|_F$ denotes the Frobenius norm, and $\alpha_s = 2^{-n} \mr{Tr}(H^{*} P_s)$ are the Pauli coefficients of $H^*$ associated with the multi-qubit Pauli-$Z$ operators $P_s = \bigotimes_{k=1}^n \sigma_z^{s_k}$.
\end{theorem}

Although the diagonal unitaries yielded by Theorem~\ref{theory:1} are compatible with quantum architectures, their exact implementation requires multi-qubit Pauli-$Z$ interactions of all orders, rendering them prohibitively error-prone on NISQ devices.
Theorem~\ref{theory:3} establishes that any diagonal Hamiltonian admits a unique decomposition into multi-body Pauli-$Z$ operators (see Sec.~\ref{sec:theorem3} in SI~\cite{SI}), and demonstrates that a single layer of $R_z$ rotations captures the zeroth- and first-order terms exactly, leaving a residual error $\epsilon_{\text{ansatz}}$ determined by the second- and higher-order coefficients.
This bound implies that systems with spectral content concentrated in low-order interactions admit highly faithful approximations.
Although Theorem~\ref{theory:3} addresses the static unitary $U^* = \ee^{\ii H^*}$, the bound extends to the time-dependent evolution operator $\ee^{\ii H^* t}$.
Consequently, $\epsilon_{\text{ansatz}}$ grows with $t$, a trend consistent with the experimental observations presented below.

\vspace{1em}
\sndtitle{Experimental setup}
All quantum experiments are executed on the superconducting processor ``Yudu'', accessed via the Quafu cloud platform~\cite{BAQIS2024Quafu}, deploying $8 \times 6$-qubit subcircuits for 3D reaction-diffusion, $32 \times 10$-qubit for spherical fluid dynamics, and $8 \times 10$-qubit for real-world ocean current simulations. 
Because the ring-structured entangling topology of the circuit ansatz maps directly onto the processor (see Sec.~\ref{sec:device} in SI~\cite{SI}), transpilation requires no SWAP gate insertions. 
Consequently, minor practical adjustments reduce the circuit depth from 12 to 9 layers for 3D reaction-diffusion systems, and from 17 to 9 for spherical fluid and ocean currents simulations, achieving a reduction of nearly 50\% in the most complex case. 
This demonstrates that the ansatz is hardware-friendly and well suited to NISQ-era execution.
The $h$ independent PQCs are mapped to spatially disjoint qubit subsets, enabling concurrent execution within a single job submission, thereby maximizing hardware throughput. 
Quantum resource allocation and processor specifications are detailed in Tab.~\ref{tab:setup} and Sec.~\ref{sec:device} in SI~\cite{SI}, respectively.

Throughout this work, we distinguish relative errors, denoted by $\varepsilon$, from absolute errors, denoted by $\epsilon$. 
A detailed error analysis is provided in Sec.~\ref{sec:error} in SI~\cite{SI}. 
We assess simulation performance using two metrics.  
The relative $L_2$ error $\varepsilon_{L2} = (\lrN{\hat{x}_k - x_{k} }^2_2)/\lrN{x_{k}}^2_2$ quantifies the deviation between the QKM prediction $\hat{x}_k$ and the reference $x_k$ at a discrete time step $k$. 
The error induced by quantum noise is evaluated using the Hellinger fidelity $F\lrr{Q, W} = (\sum_{i} \sqrt{q_i w_i})^2$, which measures the statistical overlap between the experimentally sampled distribution $Q$ and the noise-free ideal distribution $W$. 
Fidelity values close to unity indicate that the hardware execution more closely reproduces the ideal noise-free quantum state. 

\vspace{1em}
\sndtitle{Complexity analysis}
The $N$-dimensional Koopman observable space is encoded into $n=\log_2 N$ qubits. 
For each of the $h$ components in the discretized LCHS integral of Eq.~\eqref{eq:LCHS}, the quantum circuit comprises a state-preparation block and a time-evolution block. 
The gate complexity for preparing the initial state of $h$ PQCs is $\mc{C}_{\text{prep}}=\mc{O}(hnRr+hnR/2+4hn)=\mc{O}(hnRr)$.
The gate complexity for the time evolution of $h$ PQCs is $\mc{C}_{\text{evo}}=\mc{O}(hn)$. 
The total gate complexity is therefore 
$\mc{C}_{\text{qc}}=\mc{C}_{\text{prep}}+\mc{C}_{\text{evo}}=\mc{O}(hnRr)$.

We compare this cost against classical propagation of the same Koopman dynamics. 
Evaluating the matrix exponential $\ee^{ A t} u(0)$ classically incurs a cost $\mc{C}_{\text{cc}}$ that is optimally $\mc{O}(N)$ for sparse $A$, and ranges from $\mc{O}(N^2)$ to $\mc{O}(N^3)$ for dense exponentiation. 
Taking the sparse case as the conservative reference, 
the QKM enables a theoretical evolution speedup of $\mc{S}_\text{evo}=\mc{C}_{\text{cc}}/\mc{C}_{\text{evo}}=\mc{O}(2^n/(hn))$. 
Under the scaling $Rr = \mc{O}(n)$ and $h= \mc{O}(n)$ (see Sec.~\ref{sec:err_prep} and~\ref{sec:ablation_h} in SI~\cite{SI}), the resulting quantum speedup is $\mc{S}=\mc{C}_{\text{cc}}/\mc{C}_{\text{qc}}=\mc{O}({2^n}/{n^3})$. 
Standard projective measurements over $M$ shots incur a statistical error $\epsilon_{\text{meas}} = \mc{O}(1/\sqrt{M})$, yielding an end-to-end speedup of $\mc{S}_\text{total} = \mc{C}_{\text{cc}}/(M\mc{C}_{\text{qc}}) = \mc{O}(2^n \epsilon_{\text{meas}}^2/n^3)$.
This scaling indicates that a practical quantum speedup is achievable on a sufficiently large observable space of dimension $N\gtrsim\mc{O}(1/\epsilon_{\text{meas}}^2)$. 

Although quantum execution is efficient, the framework requires a one-time classical pre-processing phase to train NNs and the unitary operator parameters. 
This training stage, performed on a classical computer, optimizes these parameters to encode the nonlinear dynamics into the circuit rotation angles.

\firtitle{Data Availability}
The data presented in the figures and that support the other findings of this study will be publicly available upon its publication. 

\firtitle{Code Availability}
The source code has been deposited in QKM (https://github.com/YYgroup/QKM)~\cite{code}.

\firtitle{Acknowledgments}
This work has been supported by the National Natural Science Foundation of China (Grant Nos.~12525201, 12432010, 12588201, and 52306126), and the Beijing Natural Science Foundation (Grant No.~F261001). 

\firtitle{Author Contributions}
B.Z., Z.L., and Y.Yang conceived the theoretical idea. B.Z. and Z.L. developed the quantum Koopman method. B.Z. conducted the quantum simulations. D.A. and Z.M. contributed to the theoretical analysis. Y.Yu and X.X. provided the quantum hardware support. Y.Yang supervised the project. All authors contributed to data analysis, discussion of the results, and writing of the manuscript. 

\bibliographystyle{modified-apsrev4-2.bst}
\bibliography{main.bib}

\begin{thebibliography}{25}%
\makeatletter
\providecommand \@ifxundefined [1]{%
 \@ifx{#1\undefined}
}%
\providecommand \@ifnum [1]{%
 \ifnum #1\expandafter \@firstoftwo
 \else \expandafter \@secondoftwo
 \fi
}%
\providecommand \@ifx [1]{%
 \ifx #1\expandafter \@firstoftwo
 \else \expandafter \@secondoftwo
 \fi
}%
\providecommand \natexlab [1]{#1}%
\providecommand \enquote  [1]{``#1''}%
\providecommand \bibnamefont  [1]{#1}%
\providecommand \bibfnamefont [1]{#1}%
\providecommand \citenamefont [1]{#1}%
\providecommand \href@noop [0]{\@secondoftwo}%
\providecommand \href [0]{\begingroup \@sanitize@url \@href}%
\providecommand \@href[1]{\@@startlink{#1}\@@href}%
\providecommand \@@href[1]{\endgroup#1\@@endlink}%
\providecommand \@sanitize@url [0]{\catcode `\\12\catcode `\$12\catcode
  `\&12\catcode `\#12\catcode `\^12\catcode `\_12\catcode `\%12\relax}%
\providecommand \@@startlink[1]{}%
\providecommand \@@endlink[0]{}%
\providecommand \url  [0]{\begingroup\@sanitize@url \@url }%
\providecommand \@url [1]{\endgroup\@href {#1}{\urlprefix }}%
\providecommand \urlprefix  [0]{URL }%
\providecommand \Eprint [0]{\href }%
\providecommand \doibase [0]{https://doi.org/}%
\providecommand \selectlanguage [0]{\@gobble}%
\providecommand \bibinfo  [0]{\@secondoftwo}%
\providecommand \bibfield  [0]{\@secondoftwo}%
\providecommand \translation [1]{[#1]}%
\providecommand \BibitemOpen [0]{}%
\providecommand \bibitemStop [0]{}%
\providecommand \bibitemNoStop [0]{.\EOS\space}%
\providecommand \EOS [0]{\spacefactor3000\relax}%
\providecommand \BibitemShut  [1]{\csname bibitem#1\endcsname}%
\let\auto@bib@innerbib\@empty
%</preamble>
\bibitem [{\citenamefont {An}\ \emph {et~al.}(2023)\citenamefont {An},
  \citenamefont {Liu},\ and\ \citenamefont {Lin}}]{supp_An2023Linear}%
  \BibitemOpen
  \bibfield  {author} {\bibinfo {author} {\bibfnamefont {D.}~\bibnamefont
  {An}}, \bibinfo {author} {\bibfnamefont {J.-P.}\ \bibnamefont {Liu}},\ and\
  \bibinfo {author} {\bibfnamefont {L.}~\bibnamefont {Lin}},\ }\bibfield
  {title} {\bibinfo {title} {{\color{black}Linear combination of
  {{Hamiltonian}} Simulation for Nonunitary Dynamics with Optimal State
  Preparation Cost}},\ }\href@noop {} {\bibfield  {journal} {\bibinfo
  {journal} {Phys. Rev. Lett.}\ }\textbf {\bibinfo {volume} {131}},\ \bibinfo
  {pages} {150603} (\bibinfo {year} {2023})}\BibitemShut {NoStop}%
\bibitem [{\citenamefont {Wilcox}(1967)}]{supp_wilcox1967exp}%
  \BibitemOpen
  \bibfield  {author} {\bibinfo {author} {\bibfnamefont {R.~M.}\ \bibnamefont
  {Wilcox}},\ }\bibfield  {title} {\bibinfo {title} {{\color{black}Exponential
  operators and parameter differentiation in quantum physics}},\ }\href@noop {}
  {\bibfield  {journal} {\bibinfo  {journal} {J. Math. Phys.}\ }\textbf
  {\bibinfo {volume} {8}},\ \bibinfo {pages} {962} (\bibinfo {year}
  {1967})}\BibitemShut {NoStop}%
\bibitem [{\citenamefont {O'Donnell}(2014)}]{supp_o2014analysis}%
  \BibitemOpen
  \bibfield  {author} {\bibinfo {author} {\bibfnamefont {R.}~\bibnamefont
  {O'Donnell}},\ }\href@noop {} {\emph {\bibinfo {title} {Analysis of boolean
  functions}}}\ (\bibinfo  {publisher} {Cambridge University Press},\ \bibinfo
  {year} {2014})\BibitemShut {NoStop}%
\bibitem [{\citenamefont {Zhang}\ \emph {et~al.}(2025)\citenamefont {Zhang},
  \citenamefont {Lu}, \citenamefont {Zhao},\ and\ \citenamefont
  {Yang}}]{supp_zhang2025data}%
  \BibitemOpen
  \bibfield  {author} {\bibinfo {author} {\bibfnamefont {B.}~\bibnamefont
  {Zhang}}, \bibinfo {author} {\bibfnamefont {Z.}~\bibnamefont {Lu}}, \bibinfo
  {author} {\bibfnamefont {Y.}~\bibnamefont {Zhao}},\ and\ \bibinfo {author}
  {\bibfnamefont {Y.}~\bibnamefont {Yang}},\ }\bibfield  {title} {\bibinfo
  {title} {{\color{black}Data-driven quantum {Koopman} method for simulating
  nonlinear dynamics}},\ }\href@noop {} {\bibfield  {journal} {\bibinfo
  {journal} {preprint arXiv:2507.21890}\ } (\bibinfo {year}
  {2025})}\BibitemShut {NoStop}%
\bibitem [{\citenamefont {Ronneberger}\ \emph {et~al.}(2015)\citenamefont
  {Ronneberger}, \citenamefont {Fischer},\ and\ \citenamefont
  {Brox}}]{supp_ronneberger2015u}%
  \BibitemOpen
  \bibfield  {author} {\bibinfo {author} {\bibfnamefont {O.}~\bibnamefont
  {Ronneberger}}, \bibinfo {author} {\bibfnamefont {P.}~\bibnamefont
  {Fischer}},\ and\ \bibinfo {author} {\bibfnamefont {T.}~\bibnamefont
  {Brox}},\ }\bibfield  {title} {\bibinfo {title} {{\color{black}U-net:
  convolutional networks for biomedical image segmentation}},\ }in\ \href@noop
  {} {\emph {\bibinfo {booktitle} {Med. Image Comput. Comput.-Assist. Interv.
  (MICCAI)}}}\ (\bibinfo {organization} {Springer},\ \bibinfo {year} {2015})\
  pp.\ \bibinfo {pages} {234--241}\BibitemShut {NoStop}%
\bibitem [{\citenamefont {Li}\ \emph {et~al.}(2025)\citenamefont {Li},
  \citenamefont {Xie}, \citenamefont {Zhang}, \citenamefont {Zhang},\ and\
  \citenamefont {Zhao}}]{supp_li2025transformer}%
  \BibitemOpen
  \bibfield  {author} {\bibinfo {author} {\bibfnamefont {H.}~\bibnamefont
  {Li}}, \bibinfo {author} {\bibfnamefont {J.}~\bibnamefont {Xie}}, \bibinfo
  {author} {\bibfnamefont {C.}~\bibnamefont {Zhang}}, \bibinfo {author}
  {\bibfnamefont {Y.}~\bibnamefont {Zhang}},\ and\ \bibinfo {author}
  {\bibfnamefont {Y.}~\bibnamefont {Zhao}},\ }\bibfield  {title} {\bibinfo
  {title} {{\color{black}A transformer-based convolutional method to model
  inverse cascade in forced two-dimensional turbulence}},\ }\href@noop {}
  {\bibfield  {journal} {\bibinfo  {journal} {J. Comput. Phys.}\ }\textbf
  {\bibinfo {volume} {520}},\ \bibinfo {pages} {113475} (\bibinfo {year}
  {2025})}\BibitemShut {NoStop}%
\bibitem [{\citenamefont {Xie}\ \emph {et~al.}(2021)\citenamefont {Xie},
  \citenamefont {Wang}, \citenamefont {Yu}, \citenamefont {Anandkumar},
  \citenamefont {Alvarez},\ and\ \citenamefont {Luo}}]{supp_xie2021segformer}%
  \BibitemOpen
  \bibfield  {author} {\bibinfo {author} {\bibfnamefont {E.}~\bibnamefont
  {Xie}}, \bibinfo {author} {\bibfnamefont {W.}~\bibnamefont {Wang}}, \bibinfo
  {author} {\bibfnamefont {Z.}~\bibnamefont {Yu}}, \bibinfo {author}
  {\bibfnamefont {A.}~\bibnamefont {Anandkumar}}, \bibinfo {author}
  {\bibfnamefont {J.~M.}\ \bibnamefont {Alvarez}},\ and\ \bibinfo {author}
  {\bibfnamefont {P.}~\bibnamefont {Luo}},\ }\bibfield  {title} {\bibinfo
  {title} {{\color{black}{SegFormer}: simple and efficient design for semantic
  segmentation with transformers}},\ }\href@noop {} {\bibfield  {journal}
  {\bibinfo  {journal} {Adv. Neural Inf. Process. Syst.}\ }\textbf {\bibinfo
  {volume} {34}},\ \bibinfo {pages} {12077} (\bibinfo {year}
  {2021})}\BibitemShut {NoStop}%
\bibitem [{\citenamefont {{BAQIS}}(2024)}]{supp_BAQIS2024Quafu}%
  \BibitemOpen
  \bibfield  {author} {\bibinfo {author} {\bibnamefont {{BAQIS}}},\ }\href@noop
  {} {\bibinfo {title} {{\color{black}{Quafu} superconducting quantum
  computing}}},\ \bibinfo {howpublished}
  {\href{https://quafu-sqc.baqis.ac.cn/home}{https://quafu-sqc.baqis.ac.cn}}
  (\bibinfo {year} {2024})\BibitemShut {NoStop}%
\bibitem [{\citenamefont {Liu}\ \emph {et~al.}(2021)\citenamefont {Liu},
  \citenamefont {Kolden}, \citenamefont {Krovi}, \citenamefont {Loureiro},
  \citenamefont {Trivisa},\ and\ \citenamefont
  {Childs}}]{supp_liu2021efficient}%
  \BibitemOpen
  \bibfield  {author} {\bibinfo {author} {\bibfnamefont {J.-P.}\ \bibnamefont
  {Liu}}, \bibinfo {author} {\bibfnamefont {H.~{\O}.}\ \bibnamefont {Kolden}},
  \bibinfo {author} {\bibfnamefont {H.~K.}\ \bibnamefont {Krovi}}, \bibinfo
  {author} {\bibfnamefont {N.~F.}\ \bibnamefont {Loureiro}}, \bibinfo {author}
  {\bibfnamefont {K.}~\bibnamefont {Trivisa}},\ and\ \bibinfo {author}
  {\bibfnamefont {A.~M.}\ \bibnamefont {Childs}},\ }\bibfield  {title}
  {\bibinfo {title} {{\color{black}Efficient quantum algorithm for dissipative
  nonlinear differential equations}},\ }\href@noop {} {\bibfield  {journal}
  {\bibinfo  {journal} {Proc. Natl. Acad. Sci. U. S. A.}\ }\textbf {\bibinfo
  {volume} {118}},\ \bibinfo {pages} {e2026805118} (\bibinfo {year}
  {2021})}\BibitemShut {NoStop}%
\bibitem [{\citenamefont {Jennings}\ \emph {et~al.}(2025)\citenamefont
  {Jennings}, \citenamefont {Korzekwa}, \citenamefont {Lostaglio},
  \citenamefont {Sornborger}, \citenamefont {Subasi},\ and\ \citenamefont
  {Wang}}]{supp_jennings2025quantum}%
  \BibitemOpen
  \bibfield  {author} {\bibinfo {author} {\bibfnamefont {D.}~\bibnamefont
  {Jennings}}, \bibinfo {author} {\bibfnamefont {K.}~\bibnamefont {Korzekwa}},
  \bibinfo {author} {\bibfnamefont {M.}~\bibnamefont {Lostaglio}}, \bibinfo
  {author} {\bibfnamefont {A.~T.}\ \bibnamefont {Sornborger}}, \bibinfo
  {author} {\bibfnamefont {Y.}~\bibnamefont {Subasi}},\ and\ \bibinfo {author}
  {\bibfnamefont {G.}~\bibnamefont {Wang}},\ }\bibfield  {title} {\bibinfo
  {title} {{\color{black}Quantum algorithms for general nonlinear dynamics
  based on the {Carleman} embedding}},\ }\href@noop {} {\bibfield  {journal}
  {\bibinfo  {journal} {arXiv preprint arXiv:2509.07155}\ } (\bibinfo {year}
  {2025})}\BibitemShut {NoStop}%
\bibitem [{\citenamefont {Joseph}(2020)}]{supp_Joseph_2020}%
  \BibitemOpen
  \bibfield  {author} {\bibinfo {author} {\bibfnamefont {I.}~\bibnamefont
  {Joseph}},\ }\bibfield  {title} {\bibinfo {title}
  {{\color{black}{Koopman–von Neumann} approach to quantum simulation of
  nonlinear classical dynamics}},\ }\href@noop {} {\bibfield  {journal}
  {\bibinfo  {journal} {Phys. Rev. Res.}\ }\textbf {\bibinfo {volume} {2}},\
  \bibinfo {pages} {043102} (\bibinfo {year} {2020})}\BibitemShut {NoStop}%
\bibitem [{\citenamefont {Novikau}\ and\ \citenamefont
  {Joseph}(2025)}]{supp_Novikau2025Quantum}%
  \BibitemOpen
  \bibfield  {author} {\bibinfo {author} {\bibfnamefont {I.}~\bibnamefont
  {Novikau}}\ and\ \bibinfo {author} {\bibfnamefont {I.}~\bibnamefont
  {Joseph}},\ }\bibfield  {title} {\bibinfo {title} {{\color{black}Quantum
  algorithm for the advection-diffusion equation and the {{Koopman-von
  Neumann}} approach to nonlinear dynamical systems}},\ }\href@noop {}
  {\bibfield  {journal} {\bibinfo  {journal} {Comput. Phys. Commun.}\ }\textbf
  {\bibinfo {volume} {309}},\ \bibinfo {pages} {109498} (\bibinfo {year}
  {2025})}\BibitemShut {NoStop}%
\bibitem [{\citenamefont {Gray}\ and\ \citenamefont
  {Scott}(1983)}]{supp_gray1983autocatalytic}%
  \BibitemOpen
  \bibfield  {author} {\bibinfo {author} {\bibfnamefont {P.}~\bibnamefont
  {Gray}}\ and\ \bibinfo {author} {\bibfnamefont {S.}~\bibnamefont {Scott}},\
  }\bibfield  {title} {\bibinfo {title} {{\color{black}Autocatalytic reactions
  in the isothermal, continuous stirred tank reactor: isolas and other forms of
  multistability}},\ }\href@noop {} {\bibfield  {journal} {\bibinfo  {journal}
  {Chem. Eng. Sci.}\ }\textbf {\bibinfo {volume} {38}},\ \bibinfo {pages} {29}
  (\bibinfo {year} {1983})}\BibitemShut {NoStop}%
\bibitem [{sup()}]{supp_code}%
  \BibitemOpen
  \href@noop {} {\bibinfo {title} {{\color{black}The code is available at
  \href{https://github.com/YYgroup/QKM}{https://github.com/YYgroup/QKM}.}}}\BibitemShut
  {Stop}%
\bibitem [{\citenamefont {Galewsky}\ \emph {et~al.}(2004)\citenamefont
  {Galewsky}, \citenamefont {Scott},\ and\ \citenamefont
  {Polvani}}]{supp_galewsky2004initial}%
  \BibitemOpen
  \bibfield  {author} {\bibinfo {author} {\bibfnamefont {J.}~\bibnamefont
  {Galewsky}}, \bibinfo {author} {\bibfnamefont {R.~K.}\ \bibnamefont
  {Scott}},\ and\ \bibinfo {author} {\bibfnamefont {L.~M.}\ \bibnamefont
  {Polvani}},\ }\bibfield  {title} {\bibinfo {title} {{\color{black}An
  initial-value problem for testing numerical models of the global
  shallow-water equations}},\ }\href@noop {} {\bibfield  {journal} {\bibinfo
  {journal} {Tellus Ser. A-Dyn. Meteorol. Oceanogr.}\ }\textbf {\bibinfo
  {volume} {56}},\ \bibinfo {pages} {429} (\bibinfo {year} {2004})}\BibitemShut
  {NoStop}%
\bibitem [{\citenamefont {Burns}\ \emph {et~al.}(2020)\citenamefont {Burns},
  \citenamefont {Vasil}, \citenamefont {Oishi}, \citenamefont {Lecoanet},\ and\
  \citenamefont {Brown}}]{supp_burns2020dedalus}%
  \BibitemOpen
  \bibfield  {author} {\bibinfo {author} {\bibfnamefont {K.~J.}\ \bibnamefont
  {Burns}}, \bibinfo {author} {\bibfnamefont {G.~M.}\ \bibnamefont {Vasil}},
  \bibinfo {author} {\bibfnamefont {J.~S.}\ \bibnamefont {Oishi}}, \bibinfo
  {author} {\bibfnamefont {D.}~\bibnamefont {Lecoanet}},\ and\ \bibinfo
  {author} {\bibfnamefont {B.~P.}\ \bibnamefont {Brown}},\ }\bibfield  {title}
  {\bibinfo {title} {{\color{black}Dedalus: a flexible framework for numerical
  simulations with spectral methods}},\ }\href@noop {} {\bibfield  {journal}
  {\bibinfo  {journal} {Phys. Rev. Res.}\ }\textbf {\bibinfo {volume} {2}},\
  \bibinfo {pages} {023068} (\bibinfo {year} {2020})}\BibitemShut {NoStop}%
\bibitem [{\citenamefont {{E.U. Copernicus Marine Service
  (CMEMS)}}(2024)}]{supp_CMEMS}%
  \BibitemOpen
  \bibfield  {author} {\bibinfo {author} {\bibnamefont {{E.U. Copernicus Marine
  Service (CMEMS)}}},\ }\href@noop {} {\bibinfo {title} {{\color{black}Global
  ocean gridded {L4} sea surface heights and derived variables reprocessed 1993
  ongoing}}},\ \bibinfo {howpublished}
  {\href{https://doi.org/10.48670/moi-00148}{https://doi.org/10.48670/moi-00148}}
  (\bibinfo {year} {2024})\BibitemShut {NoStop}%
\bibitem [{\citenamefont {Hersbach}\ \emph {et~al.}(2020)\citenamefont
  {Hersbach}, \citenamefont {Bell}, \citenamefont {Berrisford}, \citenamefont
  {Hirahara}, \citenamefont {Hor{\'a}nyi}, \citenamefont {Mu{\~n}oz-Sabater},
  \citenamefont {Nicolas}, \citenamefont {Peubey}, \citenamefont {Radu},
  \citenamefont {Schepers} \emph {et~al.}}]{supp_hersbach2020era5}%
  \BibitemOpen
  \bibfield  {author} {\bibinfo {author} {\bibfnamefont {H.}~\bibnamefont
  {Hersbach}}, \bibinfo {author} {\bibfnamefont {B.}~\bibnamefont {Bell}},
  \bibinfo {author} {\bibfnamefont {P.}~\bibnamefont {Berrisford}}, \bibinfo
  {author} {\bibfnamefont {S.}~\bibnamefont {Hirahara}}, \bibinfo {author}
  {\bibfnamefont {A.}~\bibnamefont {Hor{\'a}nyi}}, \bibinfo {author}
  {\bibfnamefont {J.}~\bibnamefont {Mu{\~n}oz-Sabater}}, \bibinfo {author}
  {\bibfnamefont {J.}~\bibnamefont {Nicolas}}, \bibinfo {author} {\bibfnamefont
  {C.}~\bibnamefont {Peubey}}, \bibinfo {author} {\bibfnamefont
  {R.}~\bibnamefont {Radu}}, \bibinfo {author} {\bibfnamefont {D.}~\bibnamefont
  {Schepers}}, \emph {et~al.},\ }\bibfield  {title} {\bibinfo {title}
  {{\color{black}The {ERA5} global reanalysis}},\ }\href@noop {} {\bibfield
  {journal} {\bibinfo  {journal} {Q. J. R. Meteorol. Soc.}\ }\textbf {\bibinfo
  {volume} {146}},\ \bibinfo {pages} {1999} (\bibinfo {year}
  {2020})}\BibitemShut {NoStop}%
\bibitem [{\citenamefont {Zhang}\ \emph {et~al.}(2022)\citenamefont {Zhang},
  \citenamefont {Li},\ and\ \citenamefont {Yuan}}]{supp_zhang2022quantum}%
  \BibitemOpen
  \bibfield  {author} {\bibinfo {author} {\bibfnamefont {X.-M.}\ \bibnamefont
  {Zhang}}, \bibinfo {author} {\bibfnamefont {T.}~\bibnamefont {Li}},\ and\
  \bibinfo {author} {\bibfnamefont {X.}~\bibnamefont {Yuan}},\ }\bibfield
  {title} {\bibinfo {title} {{\color{black}Quantum state preparation with
  optimal circuit depth: implementations and applications}},\ }\href@noop {}
  {\bibfield  {journal} {\bibinfo  {journal} {Phys. Rev. Lett.}\ }\textbf
  {\bibinfo {volume} {129}},\ \bibinfo {pages} {230504} (\bibinfo {year}
  {2022})}\BibitemShut {NoStop}%
\bibitem [{\citenamefont {Sun}\ \emph {et~al.}(2023)\citenamefont {Sun},
  \citenamefont {Tian}, \citenamefont {Yang}, \citenamefont {Yuan},\ and\
  \citenamefont {Zhang}}]{supp_sun2023asymptotically}%
  \BibitemOpen
  \bibfield  {author} {\bibinfo {author} {\bibfnamefont {X.}~\bibnamefont
  {Sun}}, \bibinfo {author} {\bibfnamefont {G.}~\bibnamefont {Tian}}, \bibinfo
  {author} {\bibfnamefont {S.}~\bibnamefont {Yang}}, \bibinfo {author}
  {\bibfnamefont {P.}~\bibnamefont {Yuan}},\ and\ \bibinfo {author}
  {\bibfnamefont {S.}~\bibnamefont {Zhang}},\ }\bibfield  {title} {\bibinfo
  {title} {{\color{black}Asymptotically optimal circuit depth for quantum state
  preparation and general unitary synthesis}},\ }\href@noop {} {\bibfield
  {journal} {\bibinfo  {journal} {IEEE Trans. Comput.-Aided Des. Integr.
  Circuits Syst.}\ }\textbf {\bibinfo {volume} {42}},\ \bibinfo {pages} {3301}
  (\bibinfo {year} {2023})}\BibitemShut {NoStop}%
\bibitem [{\citenamefont {Meng}\ \emph {et~al.}(2025)\citenamefont {Meng},
  \citenamefont {Zhang}, \citenamefont {Yuan},\ and\ \citenamefont
  {Yang}}]{supp_meng2025geometric}%
  \BibitemOpen
  \bibfield  {author} {\bibinfo {author} {\bibfnamefont {Z.}~\bibnamefont
  {Meng}}, \bibinfo {author} {\bibfnamefont {X.}~\bibnamefont {Zhang}},
  \bibinfo {author} {\bibfnamefont {X.}~\bibnamefont {Yuan}},\ and\ \bibinfo
  {author} {\bibfnamefont {Y.}~\bibnamefont {Yang}},\ }\bibfield  {title}
  {\bibinfo {title} {{\color{black}Geometric encoding of turbulence for
  end-to-end quantum simulation}},\ }\href@noop {} {\bibfield  {journal}
  {\bibinfo  {journal} {arXiv preprint arXiv:2508.05346}\ } (\bibinfo {year}
  {2025})}\BibitemShut {NoStop}%
\bibitem [{\citenamefont {Meng}\ \emph {et~al.}(2026)\citenamefont {Meng},
  \citenamefont {Chen}, \citenamefont {Liu},\ and\ \citenamefont
  {He}}]{supp_meng2026toward}%
  \BibitemOpen
  \bibfield  {author} {\bibinfo {author} {\bibfnamefont {Z.}~\bibnamefont
  {Meng}}, \bibinfo {author} {\bibfnamefont {L.}~\bibnamefont {Chen}}, \bibinfo
  {author} {\bibfnamefont {J.-P.}\ \bibnamefont {Liu}},\ and\ \bibinfo {author}
  {\bibfnamefont {G.}~\bibnamefont {He}},\ }\bibfield  {title} {\bibinfo
  {title} {{\color{black}Toward end-to-end quantum simulation of rapidly
  distorted turbulence}},\ }\href@noop {} {\bibfield  {journal} {\bibinfo
  {journal} {J. Comput. Phys.}\ }\textbf {\bibinfo {volume} {558}},\ \bibinfo
  {pages} {114888} (\bibinfo {year} {2026})}\BibitemShut {NoStop}%
\bibitem [{\citenamefont {Sim}\ \emph {et~al.}(2019)\citenamefont {Sim},
  \citenamefont {Johnson},\ and\ \citenamefont
  {Aspuru-Guzik}}]{supp_sim2019express}%
  \BibitemOpen
  \bibfield  {author} {\bibinfo {author} {\bibfnamefont {S.}~\bibnamefont
  {Sim}}, \bibinfo {author} {\bibfnamefont {P.~D.}\ \bibnamefont {Johnson}},\
  and\ \bibinfo {author} {\bibfnamefont {A.}~\bibnamefont {Aspuru-Guzik}},\
  }\bibfield  {title} {\bibinfo {title} {{\color{black}Expressibility and
  entangling capability of parameterized quantum circuits for hybrid
  quantum-classical algorithms}},\ }\href@noop {} {\bibfield  {journal}
  {\bibinfo  {journal} {Adv. Quantum Technol.}\ }\textbf {\bibinfo {volume}
  {2}},\ \bibinfo {pages} {1900070} (\bibinfo {year} {2019})}\BibitemShut
  {NoStop}%
\bibitem [{\citenamefont {{\.Z}yczkowski}\ and\ \citenamefont
  {Sommers}(2005)}]{supp_zyczkowski2005average}%
  \BibitemOpen
  \bibfield  {author} {\bibinfo {author} {\bibfnamefont {K.}~\bibnamefont
  {{\.Z}yczkowski}}\ and\ \bibinfo {author} {\bibfnamefont {H.-J.}\
  \bibnamefont {Sommers}},\ }\bibfield  {title} {\bibinfo {title}
  {{\color{black}Average fidelity between random quantum states}},\ }\href@noop
  {} {\bibfield  {journal} {\bibinfo  {journal} {Phys. Rev. A}\ }\textbf
  {\bibinfo {volume} {71}},\ \bibinfo {pages} {032313} (\bibinfo {year}
  {2005})}\BibitemShut {NoStop}%
\bibitem [{\citenamefont {Goodfellow}\ \emph {et~al.}(2016)\citenamefont
  {Goodfellow}, \citenamefont {Bengio}, \citenamefont {Courville},\ and\
  \citenamefont {Bengio}}]{supp_goodfellow2016deep}%
  \BibitemOpen
  \bibfield  {author} {\bibinfo {author} {\bibfnamefont {I.}~\bibnamefont
  {Goodfellow}}, \bibinfo {author} {\bibfnamefont {Y.}~\bibnamefont {Bengio}},
  \bibinfo {author} {\bibfnamefont {A.}~\bibnamefont {Courville}},\ and\
  \bibinfo {author} {\bibfnamefont {Y.}~\bibnamefont {Bengio}},\ }\href@noop {}
  {\emph {\bibinfo {title} {Deep learning}}},\ Vol.~\bibinfo {volume} {1}\
  (\bibinfo  {publisher} {MIT press Cambridge},\ \bibinfo {year}
  {2016})\BibitemShut {NoStop}%
\end{thebibliography}

\let\addcontentsline\oldaddcontentsline

\clearpage
%\resetlinenumber
\onecolumngrid

\begin{center}
    \textbf{\large Supplementary Information for \\
    ``Quantum simulation of real-world nonlinear dynamics via Koopman method''}\\[.2cm]
    Baoyang Zhang$^{1}$, Dong An$^{2}$, Zhaoyuan Meng$^{3}$, Yefei Yu$^{4}$,
    Xiaoxiao Xiao$^{4}$, Zhen Lu$^{1,*}$ and Yue Yang$^{1,5,\dagger}$ \\[.1cm]
    {\itshape ${}^1$ State Key Laboratory for Turbulence and Complex Systems, \\
    School of Mechanics and Engineering Science, Peking University, Beijing 100871, China} \\
    {\itshape ${}^2$ Beijing International Center for Mathematical Research, \\
    Peking University, Beijing 100871, China} \\
    {\itshape ${}^3$ Institute of Mechanics, State Key Laboratory of Nonlinear Mechanics, \\
    Chinese Academy of Sciences, Beijing 100190, China} \\
    {\itshape ${}^4$ Beijing Academy of Quantum Information Sciences, Beijing 100193, China} \\
    {\itshape ${}^5$ HEDPS-CAPT, Peking University, Beijing 100871, China} \\
    {\small $^*$ zhen.lu@pku.edu.cn \qquad $^\dagger$ yyg@pku.edu.cn}
\end{center}

%---------------------------------------------------------------------------
\maketitle
\setcounter{page}{1}

\tableofcontents

% \appendix
% \counterwithout{equation}{section}
% \renewcommand{\thesubsection}{\Alph{subsection}}

\beginsupplement
\renewcommand{\thepage}{S\arabic{page}}
\renewcommand{\citenumfont}[1]{S#1}
\renewcommand{\bibnumfmt}[1]{[S#1]}

\section{Theorem proofs}

\subsection{Proof of Theorem~\ref{theory:1}}\label{sec:theorem1}
We begin by recalling the LCHS theorem~\cite{supp_An2023Linear} for time-independent matrices. For a general matrix $\tilde{ A} = \tilde{ L} + \ii\tilde{ H}$, with $\tilde{ L}=(\tilde{ A}+\tilde{ A}^\dagger)/2$, $\tilde{ H}=(\tilde{ A}-\tilde{ A}^\dagger)/2\ii$ and $\tilde{ L} \succeq 0$, the non-unitary time evolution under $-\tilde{ A}$ is given by
\begin{equation}\label{eq:original_lchs}
    \ee^{-\tilde{ A} t} = \int_{\mathbb{R}} \frac{1}{\pi(1+k^2)} \ee^{-\ii(\tilde{ H} + k\tilde{ L})t} dk.
\end{equation}
To map the original LCHS formulation to our context, we define $\tilde{ A} \coloneqq - A$, where $ A =  L + \ii H$ with Hermitian matrices $ L$ and $ H$ subject to the stability condition $ L \preceq 0$. It follows that the corresponding components are $\tilde{ L} = - L$ and $\tilde{ H} = - H$.
The condition $ L \preceq 0$ trivially guarantees that $\tilde{ L} \succeq 0$, thereby satisfying the requirement of the original theorem.
Substituting these components into Eq.~\eqref{eq:original_lchs} yields
\begin{equation}\label{eq:2_lchs}
    \ee^{ A t} = \ee^{-\tilde{ A} t} = \int_{\mathbb{R}} \frac{1}{\pi(1+k^2)} \ee^{-\ii(- H - k L)t} dk = \int_{\mathbb{R}} \frac{1}{\pi(1+k^2)} \ee^{\ii( H + k L)t} dk.
\end{equation}

Next, we define the parameterized matrices $ G(k) \coloneqq  H + k L$. Since both $ H$ and $ L$ are Hermitian, $ G(k)$ remains strictly Hermitian for all $k \in \mathbb{R}$. 
Consequently, by the spectral theorem, $ G(k)$ admits a unitary diagonalization
\begin{equation}
     G(k) =  V(k) \Lambda(k)  V^\dagger(k),
\end{equation}
where $\Lambda(k)$ is a real diagonal matrix containing the eigenvalues of $ G(k)$, and $ V(k)$ is the unitary matrix composed of its corresponding eigenvectors.
Using the standard property of matrix exponentials for diagonalizable matrices, the unitary evolution operator associated with $ G(k)$ can be expanded as
\begin{equation}\label{eq:diag}
    \ee^{\ii  G(k) t} =  V(k) \ee^{\ii \Lambda(k) t}  V^\dagger(k).
\end{equation}
Substituting Eq.~\eqref{eq:diag} into Eq.~\eqref{eq:2_lchs} directly yields the desired result
\begin{equation}
    \ee^{ A t} = \int_{\mathbb{R}} \frac{1}{\pi(1+k^2)}  V(k) \ee^{\ii \Lambda(k) t}  V^\dagger(k) dk.
\end{equation}
\qed

The condition $L \preceq 0$ is satisfied without loss of generality. For any $ A$, the substitution $ u\lrr{t}=\ee^{bt}c\lrr{t}$ with the shift parameter $b \ge \lambda_{\max}\lrr{ L}$ yields a modified system whose Hermitian part $ L - b I$ is negative semi-definite.

\subsection{Proof of Theorem~\ref{theory:2}}\label{sec:theorem2}

The continuous diagonal LCHS formulation expresses the evolution operator $\ee^{At}$ as 
\begin{equation}\label{eq:supp_lchs}
    U(t) = \ee^{At} = \int_{\mbb{R}} p(k) U_k(t) dk,
\end{equation}
where $p(k) = \frac{1}{\pi(1+k^2)}$ is the density of the Cauchy-Lorentz distribution with $\int p(k)dk=1$, and $U_k(t) = V(k)\ee^{\ii\Lambda(k)t}V^\dagger(k) = \ee^{\ii(H+kL)t}$ denotes parameterized unitary operators. 

In practice, the integral in Eq.~\eqref{eq:supp_lchs} is approximated by a discrete quadrature over $h$ points. We first truncate the infinite integration domain to a symmetric interval $[-K, K]$. Assuming a uniform sampling strategy with grid spacing $\Delta k = 2K/h$, the total approximation error $\epsilon_{\text{spec}}$ is bounded by the sum of the truncation error $\epsilon_{\text{trunc}}$ and the quadrature error $\epsilon_{\text{quad}}$. By the triangle inequality, we have
\begin{align}
    \epsilon_{\text{spec}} &\coloneqq \left\| \int_{\mbb{R}} p(k) U_k(t) dk - \sum_{j=1}^{h} \Delta k \cdot p(k_j) U_{k_j}(t) \right\|_2 \nonumber\\
    &= \left\| \int_{\mbb{R}} p(k) U_k(t) dk - \int_{-K}^{K} p(k) U_k(t) dk + \int_{-K}^{K} p(k) U_k(t) dk - \sum_{j=1}^{h} \Delta k \cdot p(k_j) U_{k_j}(t) \right\|_2 \nonumber\\
    &\le \left\| \int_{\mbb{R}} p(k) U_k(t) dk - \int_{-K}^{K} p(k) U_k(t) dk \right\|_2 + \left\| \int_{-K}^{K} p(k) U_k(t) dk - \sum_{j=1}^{h} \Delta k \cdot p(k_j) U_{k_j}(t) \right\|_2 \nonumber\\
    &= \underbrace{\left\| \int_{|k| > K} p(k) U_k(t) dk \right\|_2}_{\epsilon_{\text{trunc}}} + \underbrace{\left\| \int_{-K}^{K} p(k) U_k(t) dk - \sum_{j=1}^{h} \Delta k \cdot p(k_j) U_{k_j}(t) \right\|_2}_{\epsilon_{\text{quad}}}, \label{eq:lchs_error}
\end{align}
where $k_j=-K+j\Delta k$ denotes the $j$-th quadrature point and $k_0=-K$. 

Given that $U_k(t)$ is a unitary operator, its spectral norm is exactly $\|U_k(t)\|_2=1$. Exploiting the standard inequality $\arctan(x) \le x$ for $x \ge 0$, the truncation error $\epsilon_{\text{trunc}}$ is bounded as follows
\begin{align}
    \epsilon_{\text{trunc}} 
    &\le \int_{|k| > K} p(k) \| U_k(t) \|_2 dk \nonumber\\
    &= \frac{2}{\pi} \arctan\left(\frac{1}{K}\right) \nonumber\\
    &\le \frac{2}{\pi K}. \label{eq:trunc}
\end{align}
Define $f(k)= p(k)U_k(t)$.
The quadrature error scales as
\begin{align}
    \epsilon_{\text{quad}} 
    &= \left\| \sum_{j=1}^{h} \int_{k_{j-1}}^{k_j}  \lrs{f(k)-f(k_j)}dk \right\|_2 \nonumber\\
    &\le \sum_{j=1}^{h} \int_{k_{j-1}}^{k_j}  \left\| f(k)-f(k_j)\right\|_2 dk  \nonumber\\
    &\le \sum_{j=1}^{h} \int_{k_{j-1}}^{k_j}  | k-k_j| dk \cdot \max_{k\in [-K, K]} \left\| f^{\prime}(k)\right\|_2 \nonumber\\
    &= \frac{2K^2}{h}\max_{k\in [-K, K]}\left\|f^{\prime}(k)\right\|_2 \nonumber\\
    &\le \frac{2K^2}{h}\max_{k\in [-K, K]} \lrs{|p^\prime (k)|+|p(k)|\left\| \frac{\partial U_k(t)}{\partial k} \right\|_2} \nonumber\\
    &\le \frac{2K^2}{h}\max_{k\in [-K, K]}|p^\prime (k)| + \frac{2 K^2}{h}\max_{k\in [-K, K]}|p (k)| \cdot \max_{k\in [-K, K]}\left\| \frac{\partial U_k(t)}{\partial k} \right\|_2. 
    \label{eq:err_quad}
\end{align}
Note that
\EQ\label{eq:quad_px}
\max_{k\in [-K, K]}|p (k)|=\frac{1}{\pi} \quad \mr{and}\quad \max_{k\in [-K, K]}|p^\prime (k)|\le |p^\prime (\frac{\sqrt{3}}{3})|=\frac{3\sqrt{3}}{8\pi}.
\EN
By Duhamel’s Formula~\cite{supp_wilcox1967exp}, we have
\EQ\label{eq:quad_uk}
\left\| \frac{\partial U_k(t)}{\partial k} \right\|_2 = \left\| i \int_0^t \ee^{\ii G(k)(t-\tau)} L \ee^{\ii G(k)\tau} d\tau \right\|_2 \le t \left\| L \right\|_2 .
\EN
Substituting Eq.~\eqref{eq:quad_px} and Eq.~\eqref{eq:quad_uk} into Eq.~\eqref{eq:err_quad}, we obtain
\EQ\label{eq:quadr}
\epsilon_{\text{quad}} \le \frac{2K^2}{h} \lrr{\frac{3\sqrt{3}}{8\pi} + \frac{t \| L \|_2}{\pi}}.
\EN

Substituting Eq.~\eqref{eq:trunc} and Eq.~\eqref{eq:quadr} into Eq.~\eqref{eq:lchs_error} yields
\begin{equation}\label{eq:lchs_err2}
    \epsilon_{\text{spec}} \le \frac{2}{\pi K} + \frac{2K^2}{h} \lrr{\frac{3\sqrt{3}}{8\pi} + \frac{t \| L \|_2}{\pi}}.
\end{equation}
By optimizing with the geometric mean inequality, the overall error is bounded by
\begin{equation}\label{eq:eps_spec}
    \epsilon_{\text{spec}} \le \frac{3}{\pi}\lrr{\frac{3\sqrt{3}+8t \| L \|_2}{4h}}^{1/3},
\end{equation}
where the optimal truncation radius $K=\lrr{\frac{3\sqrt{3}+8t \| L \|_2}{4h}}^{-1/3}$ minimizes this upper bound.
\qed

\subsection{Proof of Theorem~\ref{theory:3}}\label{sec:theorem3}

Consider the functional space on the $n$-dimensional Boolean hypercube, $\mc{H}=\{f\mid f:\{0,1\}^n\to\mbb{R}\}$.
For any $f,g\in\mc{H}$, the normalized inner product is defined as the expectation over all inputs,
\begin{equation}
\langle f, g \rangle = \frac{1}{2^n} \sum_{x \in \{0,1\}^n} f(x)g(x).
\end{equation}
This functional space $\mc{H}$ is spanned by the orthonormal Walsh basis $\{w_s\}_{s \in \{0,1\}^n}$~\cite{supp_o2014analysis}, with each element
\begin{equation}
    w_s(x) = (-1)^{s \cdot x} \quad\mr{for}\quad x \in \{0,1\}^n
\end{equation}
where $s \cdot x = \sum_{k=1}^n s_k x_k \pmod 2$ represents the bitwise inner product of $s$ and $x$.

Because the action of a diagonal unitary $U^*=\ee^{\ii H^*}$ on any computational basis $\ket{x}=\ket{x_1}\otimes \cdots \otimes \ket{x_n}$ yields a specific phase distribution $U^*\ket{x}=\ee^{\ii f(x)}\ket{x}$, any diagonal Hamiltonian $H^*$ can be rigorously identified as an element of the space $\mathcal{H}$, establishing a mapping between the operator’s eigenvalues and the Boolean scalar field 
\begin{align}
H^*\ket{x} &=f(x)\ket{x} \nonumber \\ 
&= \sum_{s \in \{0,1\}^n} \alpha_s w_s(x) \ket{x} \nonumber \\ 
&= \sum_{s \in \{0,1\}^n} \alpha_s (-1)^{\sum_{k=1}^n s_k x_k} \ket{x_1}\otimes\ket{x_2}\otimes \cdots \otimes \ket{x_n} \nonumber \\
&= \sum_{s \in \{0,1\}^n} \alpha_s \lrs{(-1)^{s_1 x_1}\ket{x_1}}  \otimes  \lrs{(-1)^{s_2 x_2}\ket{x_2}} \otimes \cdots \otimes \lrs{(-1)^{s_n x_n}\ket{x_n}}  \nonumber \\
&= \sum_{s \in \{0,1\}^n} \alpha_s \lrr{\sigma_{z}^{s_1}\ket{x_1}} \otimes \lrr{\sigma_{z}^{s_2}\ket{x_2}} \cdots \otimes \lrr{\sigma_{z}^{s_n}\ket{x_n}}  \nonumber \\
&= \sum_{s \in \{0,1\}^n} \alpha_s P_s \ket{x}. \label{eq:Hstar_ori}
\end{align}
In this decomposition, $P_s = \bigotimes_{k=1}^n \sigma_z^{s_k}$ denote the multi-qubit Pauli-$Z$ operators, and the Pauli coefficients $\alpha_s$ are precisely the Walsh-Fourier transform of the eigenvalue spectrum $f(x)$, given by $\alpha_s = \langle f, w_s \rangle = 2^{-n} \mr{Tr}(H^{*} P_s)$.
From Eq.~\eqref{eq:Hstar_ori}, we obtain an isomorphism between the functional Walsh basis and the multi-qubit Pauli-$Z$ operators via the relation
\EQ\label{eq:map}
P_s \ket{x} = w_s(x) \ket{x} \quad\mr{for}\quad s,x \in \{0,1\}^n .
\EN
Since Eq.~\eqref{eq:map} holds for all computational basis, the target diagonal Hamiltonian
\EQ\label{eq:Hstar}
H^* = \sum_{s \in \{0,1\}^n} \alpha_s P_s
\EN
possesses a unique spectral decomposition, 
where the Hamming weight $|s|$ specifies the $|s|$-body interaction order, representing the number of qubits actively participating in each multi-body Pauli operator $P_s$.

We now establish the equivalent representation of the subspace $\mc{F}$ for an $n$-qubit single-layer $R_z$ circuit.
Using the mutual commutativity of the Pauli-$Z$ operators, we have
\begin{equation}\label{eq:mcF_ori}
    \ee^{\ii\phi} \bigotimes_{j=1}^n R_z(\theta_j) =  \exp\lrs{\ii \phi I_n - \sum_{j=1}^n \ii\frac{\theta_j}{2} I_1^{\otimes (j-1)} \otimes \sigma_z \otimes I_1^{\otimes (n-j)} },
\end{equation}
where $I_1$ denotes the $2\times 2$ identity matrix and $I_n$ denotes the $2^n\times 2^n$ identity matrix.
By identifying $\beta_0 = \phi$ and $\beta_s = -\theta_j/2$ for strings with Hamming weight $|s|=1$ in Eq.~\eqref{eq:mcF_ori}, the ansatz subspace can be equivalently expressed as
\EQ\label{eq:mcF}
\mc{F}=\{ \ee^{\ii H} \mid H=\beta_0 I_n + \sum_{|s|=1} \beta_s P_s,\enspace \beta_s\in\mbb{R},\enspace s\in\{0,1\}^n \}.
\EN
This equivalence demonstrates that $\mc{F}$ is the manifold of unitaries whose generators are restricted to the $0$-local and $1$-local spectral components of the functional space $\mc{H}$.

Given by the normalized Frobenius norm as the distance metric, the approximation error for any $U \in \mathcal{F}$ is
\begin{align}
    \frac{1}{2^n} \| U - U^* \|_F^2 &= \frac{1}{2^n} \mr{Tr} \lrs{\lrr{U - U^*}^\dagger \lrr{U - U^*}} \nonumber \\
    & = \mr{Tr} \lrr{U^{\dagger} U + U^{*\dagger} U^* - U^\dagger U^* - U^{*\dagger} U} \nonumber \\
    & = \frac{1}{2^n} \lrs{2^n+2^n-2\mr{Re}\lrr{U^\dagger U^*}} \nonumber \\
    & = 2 -\frac{1}{2^{n-1}}\mr{Re}\mr{Tr}\lrr{U^\dagger U^*}. \label{eq:object_ori}
\end{align}
Since all $P_s$ are diagonal matrices in the computational basis, they form a mutually commuting set $[P_s, P_{s'}] = 0$. From Eqs.~\eqref{eq:Hstar} and \eqref{eq:mcF}, this commutativity allows the unitary overlap to be expressed as
\begin{align}
    U^\dagger U^* &= \exp\lrr{-\ii \beta_0 I_n - \sum_{|s|=1} \ii\beta_s P_s}\cdot \exp\lrr{\sum_{s \in \{0,1\}^n} \ii\alpha_s P_s} \nonumber \\
    &= \exp\lrs{\ii\lrr{\alpha_0-\beta_0} I_n + \sum_{|s|=1} \ii\lrr{\alpha_s-\beta_s} P_s + \sum_{|s|\ge 2} \ii\alpha_s P_s}. \label{eq:uu}
\end{align}
Substituting Eq.~\eqref{eq:uu} into Eq.~\eqref{eq:object_ori}, and utilizing Eq.~\eqref{eq:map}, we obtain
\begin{align}
\frac{1}{2^n} \| U - U^* \|_F^2
&= 2 - \frac{1}{2^{n-1}} \sum_{x \in \{0,1\}^n} \cos \lrs{ (\alpha_0-\beta_0) + \sum_{|s|=1}(\alpha_s-\beta_s)w_s(x) + \sum_{|s|\ge 2}\alpha_s w_s(x) } \nonumber \\ 
&=\frac{1}{2^{n-1}} \sum_{x \in \{0,1\}^n} \lrc{1-\cos \lrs{ (\alpha_0-\beta_0) + \sum_{|s|=1}(\alpha_s-\beta_s)w_s(x) + \sum_{|s|\ge 2}\alpha_s w_s(x) }} \nonumber \\ 
&= \frac{4}{2^n} \sum_{x \in \{0,1\}^n} \sin^2 \lrs{ \frac{(\alpha_0-\beta_0) + \sum_{|s|=1}(\alpha_s-\beta_s)w_s(x) + \sum_{|s|\ge 2}\alpha_s w_s(x)}{2}}. \label{eq:object_simple}
\end{align}

To establish the lower bound, we apply the Taylor expansion $\sin^2(\theta) = \theta^2 + \mc{O}(\theta^4)$ and the orthonormality of the Walsh functions $\langle w_s, w_{s'} \rangle = \delta_{ss'}$ (with the Kronecker delta function $\delta_{ss'}$) to Eq.~\eqref{eq:object_simple}, yielding
\begin{align}
\frac{1}{2^n} \| U - U^* \|_F^2 &= \frac{1}{2^n} \sum_{x \in \{0,1\}^n} \lrs{ (\alpha_0-\beta_0) + \sum_{|s|=1}(\alpha_s-\beta_s)w_s(x) + \sum_{|s|\ge 2}\alpha_s w_s(x) }^2 + \mc{O}\lrr{\zeta^4} \nonumber \\
&= (\alpha_0-\beta_0)^2 + \sum_{|s|=1}(\alpha_s-\beta_s)^2 + \sum_{|s|\ge 2}\alpha_s^2 + \mc{O}\lrr{\zeta^4} \label{eq:object}
\end{align}
with $\zeta = |\alpha_0-\beta_0| + \sum_{|s|=1}|\alpha_s-\beta_s| + \sum_{|s|\ge 2}|\alpha_s|$.
To minimize Eq.~\eqref{eq:object}, the variational parameters must be chosen such that the $0$-local and $1$-local residuals vanish, i.e., $\beta_0 = \alpha_0$ and $\beta_s = \alpha_s$ for all $|s|=1$. Consequently, the minimum approximation error is given by
\begin{equation}
\min_{U \in \mc{F}} \frac{1}{2^n} \| U - U^* \|_F^2 = \sum_{|s| \ge 2} \alpha_s^2 + \mathcal{O}\left( \max_{|s| \ge 2} \alpha_s^4 \right).
\end{equation}
This confirms that the single-layer $R_z$ ansatz is intrinsically limited by the high-order Walsh components of the target Hamiltonian.
\qed

\section{QKM algorithm}\label{sec:algorithm}

Training the QKM requires joint optimization of the autoencoder and Koopman operator parameters~\cite{supp_zhang2025data}.
The training dataset comprises multiple time-series trajectories of nonlinear dynamics, each containing $T+1$ consecutive states $\{ x_k\}_{k=0}^T$ sampled at intervals $\Delta t$, where $ x_k =  x\lrr{k\Delta t}$. 
We implement the supervised learning with data organized as pairs $( x_{k}, \Delta k)$, in which $\Delta k \in \lrs{0, T}$ specifies the number of time steps to be predicted ahead. 
Given such a pair, the QKM maps $ x_k$ through the encoder $\mc{E}$, evolves the resulting quantum state for a duration $\Delta k\Delta t$ via the parameterized unitary $\mc{U}$, and reconstructs the prediction through the decoder $\mc{D}$. 
The composite loss function minimizes the normalized prediction error over all training pairs,
\begin{equation}\label{eq:loss_e}
    \mc{L} = \mbb{E}_{( x_{k}, \Delta k)} \lrs{ \frac{ \lrN{ \mc{D} \circ \mc{U}^{\Delta k \Delta t} \circ \mc{E} \lrr{ x_k} -  x_{k+\Delta k} }^2_2}{\lrN{  x_{k+\Delta k} }^2_2} }.
\end{equation}
The expectation in Eq.~\eqref{eq:loss_e} is taken over the index set $\mc{I}$ consists of two complementary subsets
\begin{equation}\label{eq:index}
    \mc{I} = \underbrace{\{(\ell, 0) \mid \ell \in \{0,\dots,T\}\}}_{\text{Zero-step transitions}}
    \cup \underbrace{\{(0, \ell) \mid \ell \in \{1,\dots,T\}\}}_{\text{Initial rollouts}}.
\end{equation}
The zero-step enforces autoencoder reconstruction fidelity for $\Delta k = 0$, while the initial rollouts enforce the predictive accuracy of the learned Koopman dynamics from the initial condition $ x\lrr{0}$. 
All variational parameters, including the NN encoder–decoder weights and the quantum circuit rotation angles, are optimized jointly on NVIDIA H100 GPUs. 
Detailed architectural specifications of the autoencoder are provided in Sec.~\ref{sec:AE}.

\begin{algorithm}[H]
\caption{Training of the QKM}
\label{alg:train}
\begin{algorithmic}[1]
\Require Dataset $\{x_k, \Delta k, x_{k+\Delta k}\}$, state preparation circuit structure
\State Set circuit configuration $n, h, R, r$ and training hyperparameters
\State Initialize classical encoder $\mc{E}_{\theta}$, decoder $\mc{D}_{\phi}$, and quantum evolution parameters $\Theta$
\For{each training epoch}
    \For{each mini-batch $(x_k, \Delta k, x_{k+\Delta k})$}
        \State Encode the input field: 
        \[
         \{P_j\}_{j=1}^h \leftarrow \mc{E}_{\theta}(x_{k})            
        \]
        \State Prepare the initial quantum state:
        \[
        \ket{u_0^j} \leftarrow P_j \quad \mr{for} \quad j=1,\ldots,h 
        \]
        \State Evolve through quantum circuits:
        \[
        \ket{u_{\Delta k}^j} \leftarrow \mc{U}(\ket{u_0^j},\Theta,\Delta k) \quad \mr{for} \quad j=1,\ldots,h
        \]
        \State Compute the probability distribution:
        \[
        u_{\Delta k}^j \leftarrow \ket{u_{\Delta k}^j} \quad \mr{for} \quad j=1,\ldots,h
        \]
        \State Decode to obtain the output field:
        \[ 
        \hat{x}_{k+\Delta k} \leftarrow \mc{D}_{\phi}(u_{\Delta k}^1,\ldots,u_{\Delta k}^h)
        \]
        \State Compute loss:
        \[ 
        \mc{L} \leftarrow \frac{\|\hat{x}_{k+\Delta k} -x_{k+\Delta k}\|^2}{\|x_{k+\Delta k}\|^2} 
        \]
        \State Backpropagate and update $\theta, \phi, \Theta$ via AdamW optimizer
    \EndFor
\EndFor
\State\Return Trained model $\mc{E}_{\theta},\mc{D}_{\phi}$ and quantum evolution parameters $\Theta$
\end{algorithmic}
\end{algorithm}

\begin{algorithm}[H]
\caption{End-to-end implementation of the QKM}
\label{alg:implem}
\begin{algorithmic}[1]
\Require Physical initial condition $x(0)$, target evolution time $t$, Measurement shot count $M$
\Require Trained model $\mc{E}_{\theta},\mc{D}_{\phi}$ and parameters $\Theta$ for circuits
\State Classical pre-processing:
\[
 \{P_j\}_{j=1}^h \leftarrow \mc{E}_{\theta}\lrs{x\lrr{0}}
\]
\State Run circuits on superconducting quantum processors:
\[
\ket{u_{t}^j} \leftarrow \text{Circuit}(P_j,\Theta,t) \quad \mr{for} \quad j=1,\ldots,h
\]
\State Perform projective measurements in the Z-basis for $M$ shots:
\[
\hat{u}_{t}^j \leftarrow \ket{u_{t}^j} \quad \mr{for} \quad j=1,\ldots,h
\]
\State Classical post-processing:
\[ 
\hat{x}(t) \leftarrow \mc{D}_{\phi}(\hat{u}_{t}^1,\ldots,\hat{u}_{t}^h)
\]
\State \Return Target-time physical fields $\hat{x}(t)$
\end{algorithmic}
\end{algorithm}

Algorithm~\ref{alg:train} outlines the training process used to learn the unitary Koopman operators and autoencoder parameters from nonlinear trajectories. Algorithm~\ref{alg:implem} specifies the hardware deployment pipeline, describing the parallel execution of the $h$ discretized spectral components and the subsequent reconstruction of the evolved physical state.

\section{Comparison with existing methods}

To situate the QKM within the current landscape of quantum algorithms for nonlinear dynamics, we provide a comparative analysis with two leading paradigms: the Carleman method~\cite{supp_liu2021efficient,supp_jennings2025quantum} and the Koopman-von Neumann (KvN) method~\cite{supp_Joseph_2020, supp_Novikau2025Quantum}, as summarized in Table~\ref{tab:comparison}. 

\begin{table*}[!ht]
	\centering
    \setlength{\tabcolsep}{8pt}
    \renewcommand{\arraystretch}{1.2}
	\caption{
    Comparison of quantum algorithms for nonlinear classical dynamics.
    The table contrasts the proposed QKM with the established Carleman method and KvN method across theoretical foundations, algorithmic primitives, and hardware requirements.
    }
	\label{tab:comparison}
	\begin{tabular}{llll}
    \hline
    Aspect & Carleman method & KvN method & Our QKM \\ \hline
    Theoretical basis & Carleman linearization & Liouville equation & Koopman theory \\
    Nonlinearity requirement & Weak & Moderate (phase space) & Moderate (physical space) \\
    Quantum primitive & QLSP  & Hamiltonian simulation & Hamiltonian simulation \\
    State preparation & Oracle required & Oracle required & End-to-end encoding \\
    Hardware requirement & Fault-tolerant & Fault-tolerant & NISQ-era \\
    Drawback & Limited convergence radius & Curse of dimensionality & Data-driven optimization required\\ \hline
	\end{tabular}
\end{table*}

Existing frameworks predominantly rely on strict analytical expansion or rigorous phase-space mapping. The Carleman method employs a Taylor-like truncation to embed nonlinear ordinary differential equations (ODEs) into an infinite-dimensional linear system. While mathematically rigorous and capable of exponential quantum speedup, its convergence fundamentally restricts the method to weakly nonlinear regimes. Although recent advances accommodate broader stable systems~\cite{supp_jennings2025quantum}, the limited convergence radius remains a primary drawback. 
The KvN method maps the classical Liouville equation to a unitary Schr\"{o}dinger-like evolution. 
However, this continuous mapping requires discretizing the phase space of all possible system states, incurring an exponential overhead due to the curse of dimensionality.
The QKM departs from these rigid analytical constraints by adopting a data-driven strategy based on Koopman theory. By jointly training the Koopman propagator and the circuit ansatz, the QKM dynamically identifies an invariant subspace to effectively capture moderately nonlinear dynamics in physical space, such as reaction-diffusion systems and fluid flows, that remain beyond the efficient reach of fixed-truncation techniques, albeit at the cost of requiring data-driven optimization.

Furthermore, the algorithmic primitives and hardware prerequisites differ fundamentally across these paradigms (Table~\ref{tab:comparison}). Algorithms based on Carleman embeddings typically rely on quantum linear system problem (QLSP) solvers, while advanced implementations of the KvN method necessitate LCHS or quantum singular value transformation~\cite{supp_Novikau2025Quantum}. Both pathways demand the block-encoding of non-unitary operators, requiring deep quantum circuits, substantial ancilla overhead, and complex state-preparation oracles that are exclusively viable on fault-tolerant quantum computers. Conversely, the QKM leverages Hamiltonian simulation primitives to circumvent these prohibitive overheads, compiling the propagator directly into an ensemble of shallow, topology-native circuits. By optimizing the state encoding within the basis transformation, the QKM eliminates the need for complex block-encoding and oracle queries. This yields an end-to-end simulation pipeline that is adaptive for NISQ and early fault-tolerant hardware.

The capability of the QKM to reach moderately nonlinear dynamics rests on an alignment between the locality structure of physical Koopman generators and the native cost structure of quantum circuits. 
A general Koopman generator on a $2^n$-dimensional observable space would require all $2^n$ Pauli-$Z$ strings and thus exponentially many parameters.
In contrast, many physical systems concentrate on low-order interactions among small subsets of observables. 
Theorem~\ref{theory:3} establishes that the single-layer $R_z$ ansatz represents zeroth- and first-order Pauli-$Z$ terms exactly, so the captured modes coincide with those favored by physics.  
The NN encoder reinforces this concentration by learning an observable basis that further shifts spectral weight onto the captured modes.
The QKM-amenable regime therefore consists of systems whose Koopman generators are few-body in the learned observable basis, allowing an exponentially large observable space to be propagated with polynomial quantum resources. 
Systems with stronger high-order couplings can be addressed by adding multi-qubit gates, at the cost of deeper circuits and reduced speedup.

\section{State preparation}\label{sec:prep}
We justify the efficient state-preparation assumption. In the hardware implementation, the state-preparation circuit has depth $\mc{O}(n)$ under the scaling $Rr=\mc{O}(n)$. Here we provide a complementary oracle-based construction showing that, for structured physical initial conditions, preparing the corresponding state does not require exponential resources.

The initial state of the linearized system is $\ket{u_0} = u_0/ \|u_0\|$ where $u_0 = [g_1(x(0));\cdots,g_N(x(0))]$. 
We start with $\ket{0}$ and apply the Hadamard gate on each qubit to obtain the uniform superposition 
\begin{equation}
\frac{1}{\sqrt{N}}\sum_{j=0}^{N-1} \ket{j}. 
\end{equation}
To proceed, we append two ancilla registers initialized with all zeros, denoted as $\ket{0}_g$ and $\ket{0}_r$. 
Here $\ket{0}_g$ is used to binary encode $g_j(x(0))$ and $\ket{0}_r$ is a single qubit for rotations. 
We then apply the operator $O_g: \ket{j} \ket{0}_g \mapsto \ket{j} \ket{g_j(x(0))/g_{\max}} $, where $g_{\max} = \max\{ |g_j(x(0))| \}$, $\ket{g_j(x(0))/g_{\max}}$ represents a binary encoding of $g_j(x(0))/g_{\max}$, and the cost of constructing $O_g$ will be discussed later. 
This gives rise to the state 
\begin{equation}
        \frac{1}{\sqrt{N}}\sum_{j=0}^{N-1} \ket{j} \ket{g_j(x(0))/g_{\max}}_g \ket{0}_r. 
\end{equation}
Applying the controlled rotation $\text{c-R}: \ket{\theta}_g\ket{0}_r \mapsto \ket{\theta}_g (\theta \ket{0}_r + \sqrt{1-|\theta|^2} \ket{1}_r )$ yields 
\begin{equation}
         \frac{1}{\sqrt{N}}\sum_{j=0}^{N-1} \ket{j} \ket{g_j(x(0))/g_{\max}}_g \left( \frac{g_j(x(0))}{g_{\max}}\ket{0}_r + \sqrt{1 - \frac{|g_j(x(0))|^2}{g_{\max}^2 }} \ket{1}_r \right). 
\end{equation}
Uncompute the register with subscript $g$ by applying $O_{g}^{\dagger}$, and we obtain 
\begin{equation}
        \frac{1}{\sqrt{N}}\sum_{j=0}^{N-1} \ket{j} \ket{ 0 }_g \left( \frac{g_j(x(0))}{g_{\max}}\ket{0}_r + \sqrt{1 - \frac{|g_j(x(0))|^2}{g_{\max}^2 }} \ket{1}_r \right) = \frac{\|u_0\|}{\sqrt{N} g_{\max} }\sum_{j=0}^{N-1} \ket{u_0} \ket{ 0 }_g  \ket{0}_r  + \ket{\perp}, 
\end{equation}
where $\ket{\perp}$ represents a possibly unnormalized state with $\ket{1}_r$ in this ancilla qubit. 
Therefore, the desired state is encoded after projecting both ancilla registers to zero, which can be achieved by measurements under the computational basis or more efficiently amplitude amplification. 
Later on we will focus on the amplitude amplification procedure as the final step. 

Now we estimate the overall gate complexity of the above approach. 
In each round of the method, we apply $O_g$ and c-R for $\mathcal{O}(1)$ times. 
We assume that $g_j(x(0))$'s are classically efficiently computable (as a function of $j$ and $x(0)$ which can be classically encoded with $\mathcal{O}(n+d)$ bits under the constant machine precision), which means that the classical cost of computing $g_j(x(0))$ is $\mathcal{O}( \text{poly}(nd) )$. 
Therefore, the quantum operation $O_g$ can be implemented with $\mathcal{O}( \text{poly}(nd) )$ complexity following the standard reversible circuit construction~\cite{nielsen2010}. 
The operation c-R and its variant have been widely used in many existing quantum algorithms such as the Harrow--Hassidim--Lloyd algorithm~\cite{harrow2009quantum} and can also be efficiently implemented with constant complexity when the number of ancilla qubits in $\ket{0}_g$ is constant, which is the case when the binary encoding precision is constant (but can be sufficiently small). 
Therefore, the number of gates required in each round of the approach is $\mathcal{O}( \text{poly}(nd) )$, so the overall complexity after amplitude amplification is 
\begin{equation}\label{eqn:proof_state_prep_complexity}
        \mathcal{O} \left( \frac{\sqrt{N} g_{\max}}{ \|u_0\| } \text{poly}(nd) \right). 
\end{equation}

At a first glance, Eq.~\eqref{eqn:proof_state_prep_complexity} appears to be on a huge scale $\sim \sqrt{N}$. 
However, the norm $\|u_0\| = \sqrt{\sum_{j=0}^{N-1} |g_j(x(0))|^2 }$ on the denominator is the square root of $N$ many terms, which also scales as $\sqrt{N}$ and leads to an overall $\mathcal{O}(\text{poly}(nd))$ complexity, under the condition that a constant portion of $g_j(x(0))$'s are on the same scale as $g_{\max}$. 
Notice that such a condition can be satisfied as long as the distribution of $g_j(x(0))$ does not concentrate too much around its peak, in the cases such as $g_{\max}/g_{\min} = \mathcal{O}(1)$ where $g_{\min} = \min_j \{|g_j(x(0))| \}$ or the median of $g_j(x(0))$ is on the same scale as $g_{\max}$. 
Therefore, this theoretical construction supports the efficient state-preparation assumption in the main text, while the practical implementation further realizes this step through the NN-generated PQC with $\mc{O}(n)$ depth.

\section{Autoencoder of QKM}\label{sec:AE}

The QKM autoencoder architecture serves as the interface between the original state space and quantum representation, learning to encode input fields into quantum circuits and reconstruct evolved fields from the measurement of circuits.

The encoder $\mc{E}$ maps the input state $ x\lrr{0}$ into $h$ sets of parameters $P_k = \{ \alpha_{j,l,\ell}, \beta_{j,l,\ell}, \gamma_{j,l,\ell} \}_{j=1, l=1, \ell=1}^{n, r, R}$ for $k = 1, \dots, h$, which define the rotation angles of the $U_3$ layers that prepare the initial quantum state on $h$ quantum circuits. 
The encoding proceeds as
\begin{align}
     y_{1} &=  \mc{R}\lrs{ x\lrr{0}}, \nonumber \\
     y_{2} &=  \mc{C}\lrc{\mc{M}( y_1) + \mc{R} \circ \mc{C}\lrs{ x\lrr{0}}}, \nonumber \\
     y_{3} &= \mc{C}\lrc{
        \mc{M}\lrr{ y_{2}} + 
        \mc{T} \circ \mc{R} \circ \mc{C}\lrs{ x\lrr{0}}
    }, \label{eq:encoder}\\
    \phi_{1}\lrr{0} &= \mc{C}\lrc{
        \mc{M}\lrr{ y_{3}} + 
        \mc{T} \circ \mc{T}
        \circ \mc{R} \circ \mc{C}\lrs{ x\lrr{0}}
    }, \nonumber\\
    \phi_{2}\lrr{0} &= \mc{C}\lrc{
        \mc{M}\lrr{ y_{4}} + 
        \mc{T} \circ \mc{T} \circ \mc{T}
        \circ \mc{R} \circ \mc{C}\lrs{ x\lrr{0}}
    },\nonumber
\end{align}
where $\mc{R}(\cdot)$, $\mc{C}(\cdot)$, and $\mc{T}(\cdot)$ denote the residual, convolution, and attention blocks, respectively; $\mc{M}(\cdot)$ represents a convolution operator with a kernel size of three and stride of two for spatial downsampling and $\circ$ denotes function composition. 
The gate parameters $\{P_k\}_{k=1}^{h}$ correspond directly to the hierarchical feature maps $\{ \phi_{1}\lrr{0}, \phi_{2}\lrr{0} \}$, where the dimensions are regulated through the convolutional channel capacity to align with the PQC configurations.

The decoder $\mc{D}$ reconstructs the evolved state from the quantum-processed observables $\phi_{1}\lrr{t}$ and $\phi_{2}\lrr{t}$, mapping the measurement outcomes of the evolved quantum states back to the state space $\mc{X}$ through the learned inverse transformation
\begin{align}
     x\lrr{t} &= \mc{Q} \circ \mc{C} \lrc{w\lrs{ y_1, \mc{P}\lrr{ z_{2}}}}, \nonumber \\    
     z_{2} &= \mc{R} \circ \mc{C} \lrc{w\lrs{ y_2, \mc{P}\lrr{ z_{3}}}}, \nonumber  \\
     z_{3} &= \mc{R} \circ \mc{C} \lrc{w\lrs{ y_3, \mc{P}\lrr{ z_{4}}}},  \label{eq:decoder} \\
     z_{4} &= \mc{R} \circ \mc{C} \lrc{w\lrs{\phi_{1}\lrr{t}, \mc{P}\lrr{ z_{5}}}}, \nonumber  \\
     z_{5} &= \phi_{2}\lrr{t}, \nonumber 
\end{align}
where $\mc{P}(\cdot)$ denotes a $3\times 3$ transposed convolution with stride two for upsampling, 
$\mc{Q}(\cdot)$ is a $1\times 1$ convolution with stride one for output refinement,  
and $w(\cdot,\cdot)$ handles tensor reshaping and channel-wise concatenation. 

\begin{figure*}
    \centering
    \includegraphics[width=\textwidth]{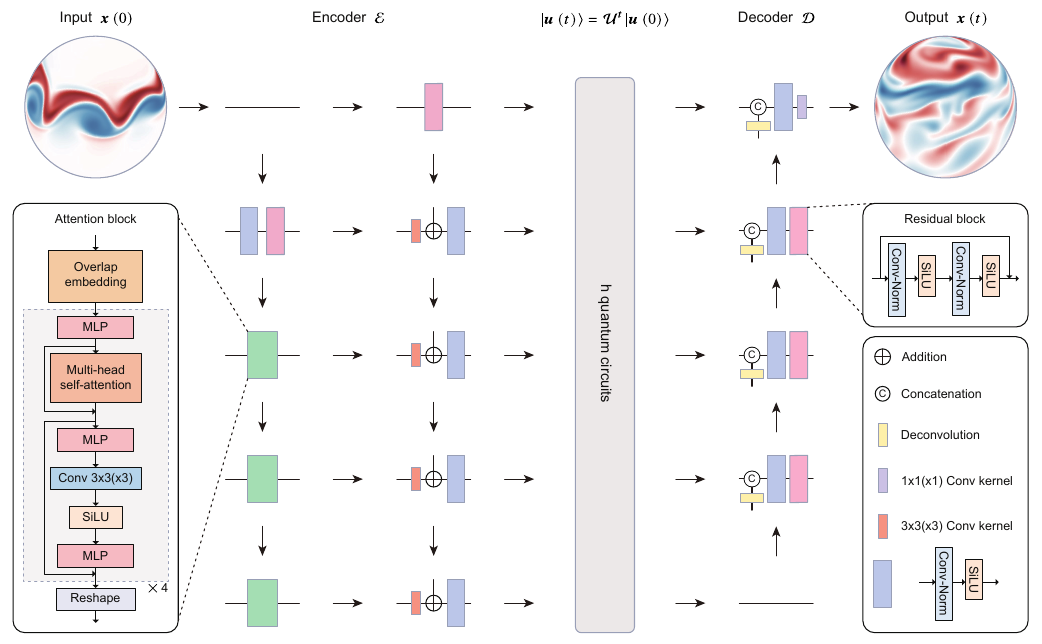}
    \caption{
    Architecture of the QKM autoencoder. 
    The encoder $\mc{E}$ transforms the physical initial condition $ x\lrr{0}$ into $h$ parameter sets to define the $U_3$ rotation angles for the initial quantum state $\ket{ u\lrr{0}}$ within each circuit. 
    The $h$ parallel quantum circuits implement the unitary evolution $\ket{ u\lrr{t}}=\mc{U}^t\ket{ u\lrr{0}}$ to represent the spectral components of the diagonalized LCHS. 
    The classical decoder $\mc{D}$ reconstructs the physical state $ x\lrr{t}$ from the measurement outcomes of the evolved quantum states $\ket{ u\lrr{t}}$. 
    The insets detail the attention blocks, residual blocks, and the computational symbol legend. 
    The attention block utilizes multi-head self-attention mechanisms and MLP-based transformations.
    The residual block architecture incorporates dual convolution and batch normalization layers with SiLU activation functions. 
    The architectural symbol legend defines computational operations including convolution, deconvolution, batch normalization, activation functions, and tensor operations.
    }
    \label{fig:net_detail}
\end{figure*}

The autoencoder is realized through a modified U-Net~\cite{supp_ronneberger2015u} architecture enhanced with vision transformer components~\cite{supp_li2025transformer}, as illustrated in Fig.~\ref{fig:net_detail}. 
The system employs three distinct computational blocks for encoding and their corresponding inverse operations for decoding, with skip connections facilitating information flow between corresponding encoder-decoder levels.
It combines convolutional neural networks for spatial processing and transformer mechanisms for long-range dependencies, forming a framework for learning the global linearized observable space of dynamical systems.

The residual blocks $\mc{R}\lrr{\cdot}$ (red block in Fig.~\ref{fig:net_detail}) integrate dual convolution (Conv)--batch normalization (BatchNorm)--sigmoid linear unit (SiLU) layers with skip connections to facilitate gradient flow and preserve fine-grained features. 
The convolution blocks $\mc{C}\lrr{\cdot}$ (blue block in Fig.~\ref{fig:net_detail}) perform spatial downsampling during encoding and feature processing during decoding. 
The attention blocks $\mc{T}\lrr{\cdot}$ (green block in Fig.~\ref{fig:net_detail}) utilize overlap embedding~\cite{supp_xie2021segformer} with 4 attention units to capture long-range spatial dependencies in the feature maps.
Each attention unit utilizes multi-head self-attention through multi-layer perceptron (MLP)-based transformations to capture spatial relationships within the encoded features. The integration of cascaded Conv--SiLU--MLP operations further facilitates channel mixing and feature transformation to improve the representational capacity of the model.

The decoder mirrors the encoder's hierarchical structure through transposed convolutions $\mathcal{P}(\cdot)$ (yellow block in Fig.~\ref{fig:net_detail}) for progressive upsampling and physical field recovery. 
The integration of time-invariant quantities $\{y_1, y_2, y_3\}$ and quantum evolution results $\{\phi_1(t), \phi_2(t)\}$ ensures the accurate reconstruction of the dynamical state. Concatenation across the latent space maintains structural integrity and enables seamless feature fusion throughout the autoencoder.

\section{Device information}\label{sec:device}

All quantum experiments were performed on the superconducting quantum processor ``Yudu''~\cite{supp_BAQIS2024Quafu}. ``Yudu'' consists of frequency-tunable transmon qubits connected by tunable couplers, with a lattice connectivity that supports nearest-neighbour two-qubit operations. The native gate set used in our experiments is composed of single-qubit rotations and controlled-$Z$ gates, summarized in Table~\ref{tab:yudu}.

\begin{table}[!ht]
\centering
\caption{
Calibration data for the qubits and $\mr{CZ}$ couplings used in the hardware experiments on the superconducting quantum processor ``Yudu''. 
Gate fidelities $F_{\mr{1Q}}$ and $F_{\mr{CZ}}$ are reported as percentages (\%), while coherence times $T_1$ and $T_2^*$ are reported in $\mu$s. 
Qubits shared by different subcircuits are listed only once. $\mr{CZ}$ labels omit the leading ``Q''. 
}
\setlength{\tabcolsep}{3.2pt}
\renewcommand{\arraystretch}{1.08}
\begin{tabular}{@{}lccc@{\hspace{0.9em}}lccc@{\hspace{0.9em}}lccc@{\hspace{1.4em}}lc@{\hspace{0.9em}}lc@{\hspace{0.9em}}lc@{}}
\toprule
\multicolumn{12}{@{}l}{Qubit properties} &
\multicolumn{6}{l}{$\mr{CZ}$ couplings} \\
\cmidrule(r){1-12}\cmidrule(l){13-18}
Qubit & $F_{\mr{1Q}}$ & $T_1$ & $T_2^*$ &
Qubit & $F_{\mr{1Q}}$ & $T_1$ & $T_2^*$ &
Qubit & $F_{\mr{1Q}}$ & $T_1$ & $T_2^*$ &
$\mr{CZ}$ & $F_{\mr{CZ}}$ & $\mr{CZ}$ & $F_{\mr{CZ}}$ & $\mr{CZ}$ & $F_{\mr{CZ}}$ \\
\midrule
Q25 & 99.93 & 44.6 & 3.8 &
Q38 & 99.93 & 57.1 & 7.6 &
Q44 & 99.95 & 54.8 & 8.2 &
25--31 & 98.77 & 37--31 & 99.22 & 43--49 & 99.50 \\

Q30 & 99.85 & 48.3 & 2.2 &
Q42 & 99.93 & 48.7 & 3.9 &
Q49 & 99.94 & 52.0 & 4.8 &
30--25 & 97.69 & 38--43 & 98.42 & 44--50 & 98.55 \\

Q31 & 99.90 & 53.0 & 10.1 &
Q43 & 99.93 & 37.8 & 14.6 &
Q50 & 99.95 & 46.3 & 7.3 &
31--38 & 99.59 & 38--44 & 99.17 & 49--42 & 98.87 \\

Q36 & 99.89 & 45.0 & 7.4 &
     &        &      &     &
     &        &      &     &
36--30 & 98.64 & 42--36 & 99.34 & 50--43 & 97.85 \\

Q37 & 99.92 & 42.8 & 6.8 &
     &        &      &     &
     &        &      &     &
42--37 & 98.99 &        &       &        &       \\
\bottomrule
\end{tabular}
\label{tab:yudu}
\end{table}

\begin{figure}
    \centering
    \includegraphics[scale=1.0]{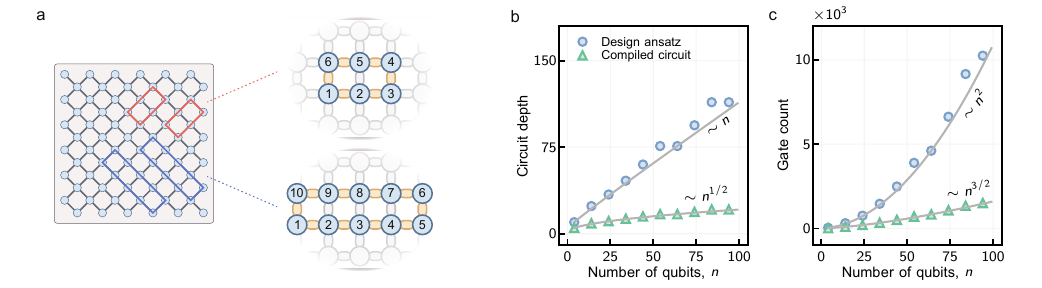}
    \caption{
    Quantum processor architecture, topology-native subcircuit mapping, and resource scaling.
    (a) Qubit layout and lattice connectivity of the superconducting quantum processor ``Yudu'', together with representative 6- and 10-qubit subcircuit mappings used for the 3D reaction--diffusion benchmark and for the spherical fluid and ocean-current benchmarks, respectively.
    The subcircuits are assigned to disjoint qubit subsets to enable parallel hardware execution.
    (b,c) Scaling of circuit depth (b) and gate count (c) with the number of qubits $n$, comparing the design ansatz before compilation with the compiled circuits after hardware mapping.
    The design ansatz exhibits approximately linear depth scaling and quadratic gate-count scaling, while topology-native compilation reduces the implemented depth and gate count. 
    All complexity analyses are based on design ansatz (pre-compilation) for fairness.
    } 
    \label{fig:chip}
\end{figure}

The QKM circuits are mapped to hardware-native subgraphs of the processor, as shown in Fig.~\ref{fig:chip}. For the 3D reaction--diffusion benchmark, each spectral component is implemented on a 6-qubit subcircuit. For the spherical fluid dynamics and real-world ocean-current benchmarks, each component is implemented on a 10-qubit subcircuit. Different spectral components are assigned to disjoint qubit subsets, enabling parallel execution on the same processor. The subcircuit topology is chosen to match the ring-like entangling layout of the QKM ansatz, thereby reducing compilation overhead and avoiding unnecessary SWAP operations. 

We also evaluate the resource scaling before and after hardware compilation. The design ansatz is used for algorithmic complexity analysis, whereas the compiled circuits quantify the practical resources required after mapping to the ``Yudu'' topology. As shown in Figs.~\ref{fig:chip}b,c, topology-native compilation reduces both circuit depth and gate count while preserving the intended circuit structure.

\section{Description of benchmarks}\label{sec:datasets}

\subsection{3D reaction-diffusion systems}\label{sec:cube}

The Gray-Scott equation~\cite{supp_gray1983autocatalytic}
\begin{align}
    \frac{\partial u}{\partial t} &= D_u \nabla^2 u - u v^2 + F(1 - u), \label{eq:gray_scott_A} \\
    \frac{\partial v}{\partial t} &= D_B \nabla^2 v + u v^2 - (F + K)v \label{eq:gray_scott_B}
\end{align}
provides a quintessential representation of complex pattern formation in reaction-diffusion systems. 
This system describes the interaction between two chemical species with concentrations $u$ and $v$, where the respective diffusion coefficients are $D_u = 2\times 10^{-5}$ and $D_v = 1\times 10^{-5}$. The feed rate $F$ and kill rate $K$ collectively determine the morphological evolution of the system, governing the transition between stationary states and complex Turing patterns.

The dataset was generated via direct numerical simulation (DNS) using the spectral method~\cite{supp_code}, with parameters set to $F=0.018$ and $K=0.051$. The computational domain spans $[-1,1]^3$ under periodic boundary conditions, discretized with $64^3$ grid points. 
We generated a dataset comprising $1200$ independent trajectories initiated from random Gaussian fields. Each trajectory consists of $61$ temporal snapshots, corresponding to a final index $T=60$ as defined in Eq.~\eqref{eq:index}, with a sampling interval of $\Delta t = 5$ covering the total evolution period $t \in [0, 300]$.
The datasets were partitioned into training, validation, and test sets with a ratio of 8:1:1. 
Model parameters were optimized using the training set, employing a cosine decay learning rate schedule with logarithmic warm-up, while the validation set guided early stopping. 
Final performance was assessed on the test set. 

\subsection{Spherical fluid dynamics}\label{sec:sphere}

Spherical shallow-water equations are instrumental in modeling global atmospheric circulation and geophysical flows. 
We adopted the barotropic instability test case ~\cite{supp_galewsky2004initial}, which involves a potent zonal jet subject to small perturbations that trigger complex vortex shedding and turbulent cascades. The governing equations 
\begin{align}
    \frac{\partial  u}{\partial t} &= -( u \cdot \nabla)  u -  \nu \nabla^4  u - g \nabla h - 2\mathbf{\Omega} \times  u, \label{eq:swe_u} \\
    \frac{\partial h}{\partial t} &= -\nabla \cdot (h u) - \nu \nabla^4 h - H \nabla \cdot  u \label{eq:swe_h}
\end{align}
are formulated in spherical coordinates to account for the Coriolis effect and the intrinsic curvature of the Earth. 
Physical parameters are scaled to Earth's characteristics, where the unit length corresponds to the mean planetary radius $6.37 \times 10^6$ meters and the temporal unit is one hour.
In this dimensionless representation, the planetary radius is $R = 1$ and the magnitude of the rotation rate $\lrn{\mathbf{\Omega}}$ is $0.26$. The angular velocity vector $\mathbf{\Omega}$ is oriented parallel to the planetary rotation axis, which determines the meridional variation of the Coriolis effect. The gravitational acceleration $g$ and the mean fluid depth $H$ are set to $19.95$ and $1.57 \times 10^{-3}$ respectively, while the hyperviscosity coefficient $\nu$ is approximately $8.66 \times 10^{-9}$ to ensure stability. 
The system is discretized using a spherical harmonic basis with $512 \times 256$ grid points across the longitudinal and latitudinal dimensions.
To mitigate singularities at the poles, the latitudinal domain is extended through the concatenation of the original grid with a version flipped meridionally and shifted by 180 degrees zonally. The resulting $512 \times 512$ grid representation ensures numerical continuity across the polar axes and preserves the physical integrity of the spherical manifold during feature extraction.

The dataset was generated through DNS using a pseudo-spectral code~\cite{supp_burns2020dedalus}.
The initial velocity field is constructed as a mid-latitude jet with a random peak velocity. To ensure physical consistency, the initial height field is obtained by solving a linear boundary value problem to satisfy the geostrophic balance condition. This stable equilibrium is subsequently perturbed by a localized Gaussian height anomaly with a random amplitude to trigger the barotropic instability.

We generated a dataset comprising $1200$ independent trajectories. Each trajectory consists of $51$ temporal snapshots, corresponding to a final index $T=50$ as defined in Eq.~\eqref{eq:index}, with a sampling interval of $\Delta t = 2$ covering the fully developed turbulent regime $t \in [400, 500]$. The resulting data were partitioned into training, validation, and test sets according to an 8:1:1 ratio. Model parameters were optimized using the training set with a cosine decay learning rate schedule preceded by a logarithmic warm-up phase. The validation set was employed to guide the early stopping criterion to prevent overfitting, while the final predictive performance was assessed on the independent test set.

\subsection{Real-world ocean currents}\label{sec:square}

Real-world ocean currents exhibit complex nonlinear dynamics, motivating the selection of the Gulf Stream as a representative benchmark to validate the performance of the QKM. The spatial domain covers latitudes from $20^{\circ}$N to $52^{\circ}$N and longitudes from $33^{\circ}$W to $65^{\circ}$W, discretized into $256 \times 256$ grid points with a spatial resolution of $0.125^{\circ}$. 
We utilize absolute geostrophic velocity fields derived from Level-4 satellite reanalysis products to represent the surface flow field~\cite{supp_CMEMS}. To incorporate geographical constraints and boundary conditions, these velocity fields are integrated with static environmental variables~\cite{supp_hersbach2020era5}, specifically the land-sea mask and the geopotential representing the regional topographic height.

Spanning the period from 1993 to 2024, the dataset is reorganized into independent trajectories with a final index $T=12$ and a 13-day temporal duration. The resulting data are partitioned into disjoint sets for training, validation, and independent testing.
Model optimization is conducted using the training sequences from 1993 to 2022 with a cosine decay learning rate schedule preceded by a logarithmic warm-up phase. The validation data from 2022 serve to guide the early stopping criterion to prevent overfitting, while the final predictive robustness is assessed on the independent test set from 2024. 

\section{Detailed error analysis}\label{sec:error}
Throughout this work, $\varepsilon$ denotes relative errors and $\epsilon$ denotes absolute errors. 
The cumulative error $\epsilon$ of the QKM is partitioned into theoretical and experimental constituents. 
\begin{align}
\epsilon &= \epsilon_{\text{th}} + \epsilon_{\text{exp}}, \nonumber\\  
\epsilon_{\text{th}} &= \epsilon_{\text{proj}} + \epsilon_{\text{spec}} + \epsilon_{\text{ansatz}}, \label{eq:error}\\ 
\epsilon_{\text{exp}} &= \epsilon_{\text{meas}}+\epsilon_{\text{prep}}+\epsilon_{\text{opt}}+\epsilon_{\text{noise}}.
\nonumber
\end{align}
Each term corresponds to a stage of the simulation pipeline. 

In the theoretical error $\epsilon_{\text{th}}$, the projection error $\epsilon_{\text{proj}}$ arises from truncating the infinite-dimensional Koopman operator to a finite $N$-dimensional subspace, which depends on the expressivity of the learned observable functions.  
The spectral sampling error $\epsilon_{\text{spec}}$ originates from approximating the integral in Eq.~\eqref{eq:LCHS} by a discrete sum over $h$ points. 
The ansatz approximation error $\epsilon_{\text{ansatz}}$ reflects the structural bias introduced by representing each diagonalized evolution operator with a single $R_z$ layer. 
This design prioritizes circuit shallowness over universal expressivity, limiting the fidelity of the unitary synthesis to the range achievable by the chosen parametric form.

In the experimental error $\epsilon_{\text{exp}}$, the measurement error $\epsilon_{\text{meas}}$ originates from the reconstruction of the state probabilities $|\langle i|u(t)\rangle|^2$ ($i=1,\ldots,N$) based on outcomes from the $h$ PQCs.
The state preparation error $\epsilon_{\text{prep}}$ stems from the $\{U_3,\mathrm{CZ}\}$ PQCs used to encode the initial state $\ket{ u\lrr{0}}$.
The optimization error $\epsilon_{\text{opt}}$ includes the convergence gap of the variational optimization and the generalization error.
The hardware noise error $\epsilon_{\text{noise}}$ encapsulates the inaccuracies introduced by gate infidelities, decoherence, and measurement errors during execution on physical quantum processors. 

While the total error $\epsilon$ and hardware noise $\epsilon_{\text{noise}}$ are empirically evaluated in the Results section and the bounds for $\epsilon_{\text{spec}}$ and $\epsilon_{\text{ansatz}}$ are provided by Theorems~\ref{theory:2} and \ref{theory:3}, respectively, this section offers analysis of the remaining terms. Specifically, we present theoretical estimation for $\epsilon_{\text{meas}}$, and quantitative discussion of $\epsilon_{\text{prep}}$ and $\epsilon_{\text{opt}}$ based on empirical observations.
Complementing the theoretical analyses in Theorems~\ref{theory:2} and \ref{theory:3}, we further provide an overall quantitative estimation of $\epsilon_{\text{th}}$. 

\subsection{Theoretical estimation of measurement error}\label{sec:err_meas}

The decoding stage of the QKM framework requires reconstructing the probability distribution $u$ with components $u_i = |\langle i|u(t)\rangle|^2$ from measurements. The reconstructed $u$ is not derived from a single quantum state, but rather from a linear combination of the outputs from the $h$ parallel circuits, in accordance with the discretized LCHS integral. For each of the $h$ circuits, standard projective measurements in the computational basis are performed with $M$ shots, leading to a statistical reconstruction error.

Let the exact target probability distribution vector of the $j$-th circuit be $u^{(j)} \in \mathbb{R}^N$ (for $j = 1, \dots, h$), where its $i$-th component $u^{(j)}_i = p_i^{(j)}$ represents the theoretical probability of observing the computational basis state $\ket{i}$, satisfying $\sum_{i=1}^N p_i^{(j)} = 1$. The overall reconstructed state vector is defined by the weighted sum $u = \sum_{j=1}^h w_j u^{(j)}$. To ensure $u$ is a valid probability distribution, the weights $w_j > 0$ are strictly positive and normalized $\sum_{j=1}^h w_j = 1$. 
The expected absolute error in the $L_2$ norm, which defines $\epsilon_{\text{meas}}$, follows directly from Jensen's inequality
\begin{equation}\label{eq:meas_1}
    \epsilon_{\text{meas}} \coloneqq \mathbb{E} \left[ \| \hat{u} - u \|_2 \right] \le \sqrt{\mathbb{E}\left[ \| \hat{u} - u \|_2^2 \right]} = \sqrt{\sum_{i=1}^{N} \text{Var}(\hat{u}_i)}.
\end{equation}
The combined empirical estimator for the $i$-th component of the total state is $\hat{u}_i = \sum_{j=1}^h w_j \hat{u}^{(j)}_i$. Because the $h$ parallel circuits are executed and measured independently, the variance of the combined estimator is the weighted sum of the individual variances
\begin{equation}\label{eq:meas_2}
    \mr{Var}(\hat{u}_i) = \sum_{j=1}^h w_j^2 \mr{Var}(\hat{u}^{(j)}_i).
\end{equation}
A single shot collapses the $n$-qubit circuit into one of the $N = 2^n$ possible computational basis states. Therefore, measuring the circuit $M$ times yields an empirical probability distribution that follows a multinomial distribution. The empirical estimator $\hat{u}^{(j)}_i$, as the observed frequency of state $\ket{i}$, is an unbiased estimator of the true probability $\mathbb{E}[\hat{u}^{(j)}_i] = p_i^{(j)} = u^{(j)}_i$, and its variance is given by
\begin{equation}\label{eq:meas_3}
    \text{Var}(\hat{u}^{(j)}_i) = \frac{p_i^{(j)}(1-p_i^{(j)})}{M}.
\end{equation}
Substituting Eq.~\eqref{eq:meas_2} and Eq.~\eqref{eq:meas_3} into Eq.~\eqref{eq:meas_1} yields an upper bound
\EQ\label{eq:meas_4}
\epsilon_{\text{meas}} \le \sqrt{\sum_{j=1}^{h}\sum_{i=1}^{N}w_j^2\frac{p_i^{(j)}(1-p_i^{(j)})}{M}}\le\sqrt{\sum_{j=1}^{h}\sum_{i=1}^{N}w_j^2\frac{p_i^{(j)}}{M}}=\sqrt{\frac{1}{M}\sum_{j=1}^{h}w_j^2}.
\EN

For the uniform LCHS quadrature over the truncated interval $[-K, K]$, the grid spacing is $\Delta k = 2K/h$, and the Cauchy-Lorentz density is universally bounded by $p(k) \le 1/\pi$. Therefore, the weight is bounded by $w_j \approx p(k_j)\Delta k_j \lesssim 2K/(\pi h)$. With the optimal $K=\mc{O}(h^{1/3})$ for $\epsilon_{\text{spec}}$ in Eq.~\eqref{eq:eps_spec}, we obtain
\EQ\label{eq:meas_5}
\sum_{j=1}^{h} w_j^2 \le \left( \max_{1 \le j \le h} w_j \right) \sum_{j=1}^{h} w_j = \max_{1 \le j \le h} w_j=\mc{O}\lrr{h^{-\frac{2}{3}}}.
\EN
Substituting Eq.~\eqref{eq:meas_5} into Eq.~\eqref{eq:meas_4} yields
\EQ\label{eq:meas_6}
\epsilon_{\text{meas}} \le \sqrt{\frac{\max_{1 \le j \le h} w_j}{M}} = \mc{O}\lrr{ M^{-\frac{1}{2}} h^{-\frac{1}{3}} }.
\EN

\subsection{Quantitative analysis of state preparation error}\label{sec:err_prep}

In the absence of ancillary qubits, preparing an arbitrary $n$-qubit quantum state requires an exponential circuit depth~\cite{supp_zhang2022quantum}, with the optimal scaling $\mc{O}(2^n/n)$~\cite{supp_sun2023asymptotically}. 
Such complexity restricts high-dimensional quantum simulations to theoretical frameworks, rather than real-device execution. 
By exploiting the intrinsic regularities of real-world physical systems, such as continuity~\cite{supp_meng2025geometric,supp_meng2026toward}, our approach effectively circumvents the exponential circuit depth typically required for arbitrary quantum state synthesis.
The state preparation error $\epsilon_{\text{prep}}$ in Eq.~\eqref{eq:error} is fundamentally governed by the expressibility of the PQCs used to encode the initial state $\ket{u(0)}$. 
To quantify the resulting representational capacity of our topology-native $\{U_3, \mathrm{CZ}\}$ ansatz, we evaluate its expressibility by comparing the ensemble of states it generates against the ensemble of Haar-random states~\cite{supp_sim2019express}. 

The evaluation relies on the distribution of state fidelities $F = |\langle \psi_{\theta} | \psi_{\phi} \rangle|^2$, obtained by uniformly sampling parameter sets $\theta$ and $\phi$ to generate state pairs.
A perfectly expressive PQC is characterized by a fidelity distribution that replicates the Haar-random ensemble~\cite{supp_sim2019express}, whose analytical probability density function is given by~\cite{supp_zyczkowski2005average}
\begin{equation}
    P_{\text{Haar}}(F) = (N-1)(1-F)^{N-2} 
\end{equation}
with $N = 2^n$. 
The expressibility is quantified by the Kullback-Leibler (KL) divergence between the empirical fidelity distribution $P_{\text{PQC}}(F)$ and the reference Haar-random distribution $P_{\text{Haar}}(F)$ , evaluated as a discrete estimator partitioned into $N_{\text{bins}}$ uniformly spaced bins
\begin{equation} \label{eq:kl_div}
    D_{\text{KL}}(P_{\text{PQC}} \| P_{\text{Haar}}) = \sum_{i=1}^{N_{\text{bins}}} \hat{P}_{\text{PQC}}(F_i) \ln \left( \frac{\hat{P}_{\text{PQC}}(F_i)}{\hat{P}_{\text{Haar}}(F_i)} \right).
\end{equation}
Under this statistical framework, a lower $D_{\text{KL}}$ value signifies a more expressive circuit that more effectively explores the Hilbert space~\cite{supp_sim2019express}.
Since the representational capacity of PQCs fundamentally constrains the ability to approximate target states, superior expressivity implies a systematic reduction in the state preparation error $\epsilon_{\text{prep}}$. Consequently, the KL divergence $D_{\text{KL}}$ serves as an empirical proxy to estimate $\epsilon_{\text{prep}}$ within the QKM.

\begin{figure*}
    \centering
    \includegraphics[width=\textwidth]{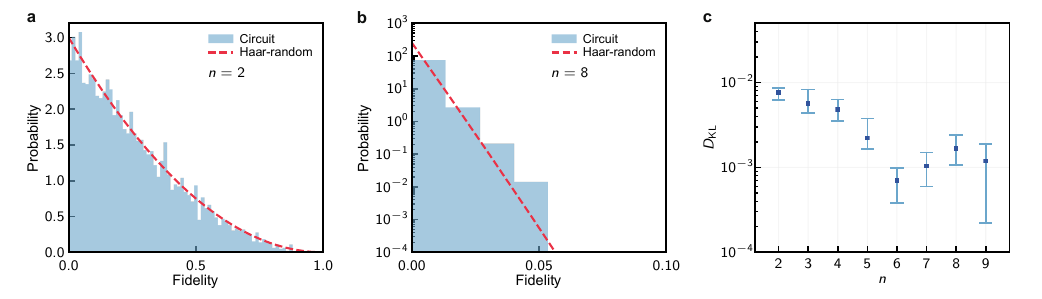}
    \caption{Expressibility analysis of the state preparation ansatz. Empirical fidelity distributions (blue histograms) generated by the topology-native $\{U_3, \mathrm{CZ}\}$ circuits for (a) $n=2$ and (b) $n=8$ qubits, compared against the analytical Haar-random probability density function (red dashed lines). The vertical axis in (b) is shown on a logarithmic scale to account for the concentration of measure in high-dimensional Hilbert spaces. (c) KL divergence $D_{\text{KL}}$ between the circuit-generated fidelity distribution and the Haar-random ensemble as a function of qubit number $n$. Markers represent the mean $D_{\text{KL}}$ value obtained from 5 independent numerical experiments, with error bars denoting the maximum and minimum values. The consistently low $D_{\text{KL}}$ values indicate that the circuit effectively generates states representative of the Hilbert space, with minimal information loss compared to a Haar-random distribution.}
    \label{fig:err_prep}
\end{figure*}

Here we configure the depth parameters $R$ and $r$ as $R = r = \lceil \sqrt{n} \rceil$. This ensures that the total circuit depth scales as $Rr \approx n$, providing a necessary balance between the circuit's representational capacity and the coherence constraints of near-term quantum hardware. 
For a given $n$, we uniformly sample pairs of parameter vectors to generate the corresponding quantum states and compute their pairwise fidelities. 

Because the empirical estimation of the KL divergence is affected by finite-sampling effects and histogram resolution, we set the sample size to $N_{\text{samples}} = 5000$ and discretize the continuous fidelity range $[0, 1]$ into $N_{\text{bins}} = 75$ uniformly spaced bins. These parameters are chosen to match the benchmarking in Ref.~\cite{supp_sim2019express}, ensuring a consistent and valid evaluation.
Fig.~\ref{fig:err_prep}(a) and Fig.~\ref{fig:err_prep}(b) illustrate the resulting fidelity distributions for $n=2$ and $n=8$ qubits, respectively. The empirical histograms generated by our topology-native circuit exhibit alignment with the analytical Haar-random probability density functions. For the larger system size ($n=8$), the fidelity distribution heavily concentrates near zero, reflecting the concentration of measure typical of high-dimensional Hilbert spaces. 
In this regime, the circuit accurately reproduces the theoretical Haar distribution over a logarithmic scale, indicating that the parameterization explores the state space uniformly and with minimal bias.
A systematic quantitative assessment of this expressibility across system sizes from $n=2$ to $n=9$ is presented in Fig.~\ref{fig:err_prep}(c). The calculated $D_{\text{KL}}$ values remain consistently low, generally on the order of $10^{-2}$ to $10^{-4}$, across all evaluated qubit counts.

For $n=10$, we the concentration of measure in high-dimensional Hilbert spaces requires finer sampling. 
Using $N_{\text{samples}}=50000$ and $N_{\text{bins}} = 750$, we obtain $D_{\text{KL}}=1.04\times 10^{-4}$. 
The KL divergence does not exhibit growth as the system scales; instead, it maintains a stable, low magnitude. 
This confirms that the representational capacity of the circuit does not degrade for larger systems, ensuring that the PQC can generate states highly representative of the full Hilbert space.

By evaluating Eq.~\eqref{eq:kl_div} across varying qubit scales $n$, we numerically demonstrate that the adopted scaling of $Rr =\mathcal{O}(n)$ provides sufficient expressibility to represent the physical initial conditions, thereby establishing an empirical estimate on $\epsilon_{\text{prep}}$ while circumventing the exponential depth typically required for arbitrary state synthesis.

\subsection{Quantitative analysis of theoretical and optimization errors}\label{sec:err_opt}

Within the QKM framework, the pre-training phase involves the joint optimization of the neural network encoder-decoder and the unitary Koopman operators. The optimization error $\epsilon_{\text{opt}}$ in Eq.~\eqref{eq:error} encapsulates the inaccuracies introduced during this variational training process. 
To quantify this constituent within the context of physical prediction, we assess it using the relative metric $\mr{relative}\;L_2\text{-}\epsilon_{\text{opt}}$, which provides a direct evaluation of the relative deviation arising from optimization.

Because our composite loss function $\mc{L}$ defined in Eq.~\eqref{eq:loss_e} directly computes the expectation of the relative $L_2$ error, the optimization error $\mr{relative}\;L_2\text{-}\epsilon_{\text{opt}}$ shares the same scale and physical interpretation as $\mr{relative}\;L_2\text{-}\epsilon$.
Following statistical learning theory~\cite{supp_goodfellow2016deep}, the expected error can be decomposed into the empirical training risk and the generalization gap.
Therefore, without the need for scaling factors, we estimate the relative theoretical and optimization error as
\begin{align}
\varepsilon_{\text{th}}\coloneqq\mr{relative}\;L_2\text{-}\epsilon_{\text{th}} &\approx\mathcal{L}_{\text{train}}\approx \ell_{\text{train}} ,\label{eq:th_error}\\
\varepsilon_{\text{opt}}\coloneqq\mr{relative}\;L_2\text{-}\epsilon_{\text{opt}} &\approx |\mathcal{L}_{\text{val}} - \mathcal{L}_{\text{train}}| \approx |\ell_{\text{val}}-\ell_{\text{train}}|, \label{eq:opt_error}
\end{align}
where $\mathcal{L}_{\text{train}}$ and $\mathcal{L}_{\text{val}}$ represent the asymptotic values of the loss function evaluated on the training and unseen validation datasets, respectively, while $\ell_{\text{train}}$ and $\ell_{\text{val}}$ denote the terminal training and validation losses achieved at the conclusion of the training process. 
 
\begin{figure*}
    \centering
    \includegraphics[width=\textwidth]{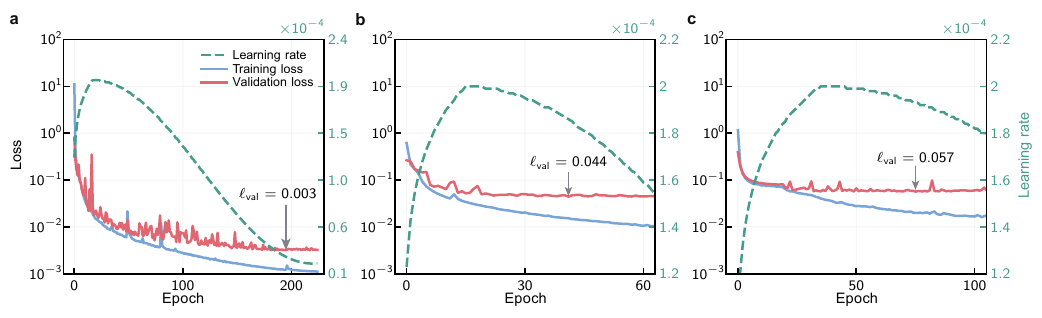}
    \caption{Learning dynamics and loss of the QKM framework. The temporal evolution of the composite loss function $\mc{L}$ and the learning rate schedule for (a) the 3D reaction-diffusion system, (b) the spherical fluid dynamics, and (c) the real-world ocean currents. In each panel, the training loss $\ell_{\text{train}}$ (blue solid line) and validation loss $\ell_{\text{val}}$ (red solid line) are plotted against the left axis on a logarithmic scale. The learning rate schedule (green dashed line), featuring a warm-up phase followed by a cosine decay, is shown on the right axis. The annotated $\ell_{\text{val}}$ indicates the terminal validation error achieved at the conclusion of training via an early stopping criterion to prevent overfitting. The minimal gap $|\ell_{\text{val}}-\ell_{\text{train}}|$ between the training and validation curves indicates that the relative optimization error $\mr{relative}\;L_2\text{-}\epsilon_{\text{opt}}$ is small, while the terminal training loss $\ell_{\text{train}}$ quantitatively estimates the theoretical error $\mr{relative}\;L_2\text{-}\epsilon_{\text{th}}$.
    }
    \label{fig:err_opt}
\end{figure*}

Figure~\ref{fig:err_opt} monitors the training and validation losses across the three benchmarks. For the 3D reaction-diffusion system (Fig.~\ref{fig:err_opt}a), the training terminates at epoch 195 with $\ell_{\text{train}} = 0.001$ and $\ell_{\text{val}} = 0.003$. For the spherical fluid dynamics (Fig.~\ref{fig:err_opt}b), the process ends at epoch 41, yielding $\ell_{\text{train}} = 0.015$ and $\ell_{\text{val}} = 0.044$. The real-world ocean current benchmark (Fig.~\ref{fig:err_opt}c) concludes at epoch 75 with $\ell_{\text{train}} = 0.020$ and $\ell_{\text{val}} = 0.057$. 
In all three cases, the validation loss closely tracks the training curve without divergence, resulting in a minimal generalization gap $|\ell_{\text{val}}-\ell_{\text{train}}|$ throughout the optimization.

Substituting these terminal values into Eqs.~\eqref{eq:th_error} and \eqref{eq:opt_error} provides quantitative measures for both $\varepsilon_{\text{th}}$ and $\varepsilon_{\text{opt}}$. Across all benchmarks, the minimal generalization gap confirms that the optimization error $\epsilon_{\text{opt}}$ remains a subdominant factor. Instead, the empirical training loss $\ell_{\text{train}}$ serves as the experimental manifestation of the cumulative theoretical error $\epsilon_{\text{th}}$, capturing the fundamental constraints imposed by the Koopman projection $\epsilon_{\text{proj}}$, the spectral sampling bias $\epsilon_{\text{spec}}$, and the ansatz structural bias $\epsilon_{\text{ansatz}}$. 

Comparing these estimates with the total $\mr{relative}\;L_2\text{-}\epsilon$ trajectories reported in the Results section allows us to distinguish the primary error sources across different regimes. For the 3D reaction-diffusion case, the minimal $\ell_{\text{train}}$ indicates that $\epsilon_{\text{th}}$ is well-suppressed, leaving the hardware noise $\epsilon_{\text{noise}}$ as the primary source of the total simulation error. Conversely, the higher $\ell_{\text{train}}$ plateaus observed in the spherical and ocean current benchmarks demonstrate that the underlying theoretical complexity, specifically the multi-scale dynamics, dominates the error budget. In these regimes, the theoretical limit $\epsilon_{\text{th}}$ becomes the primary constraint on fidelity, rendering the contribution of $\epsilon_{\text{noise}}$ a secondary factor. 
Furthermore, the statistical measurement error $\epsilon_{\text{meas}}$ can be systematically suppressed by increasing the sampling shot count $M$.
Consequently, improving simulation accuracy requires a dual approach: mitigating hardware noise for weakly nonlinear systems, and refining the theoretical representation and training convergence for complex, multi-scale dynamics.

\section{Ideal noiseless simulation}\label{sec:sim}

This section provides ideal noiseless simulations to serve as a reference for the hardware experiments discussed in the main text. By employing classical emulation of the QKM, we isolate the algorithmic performance from measurement errors $\epsilon_{\text{meas}}$ and hardware-induced noise $\epsilon_{\text{noise}}$.

\subsection{3D reaction-diffusion systems}\label{sec:sim_3D}

\begin{figure*}
    \centering
    \includegraphics[scale=1.0]{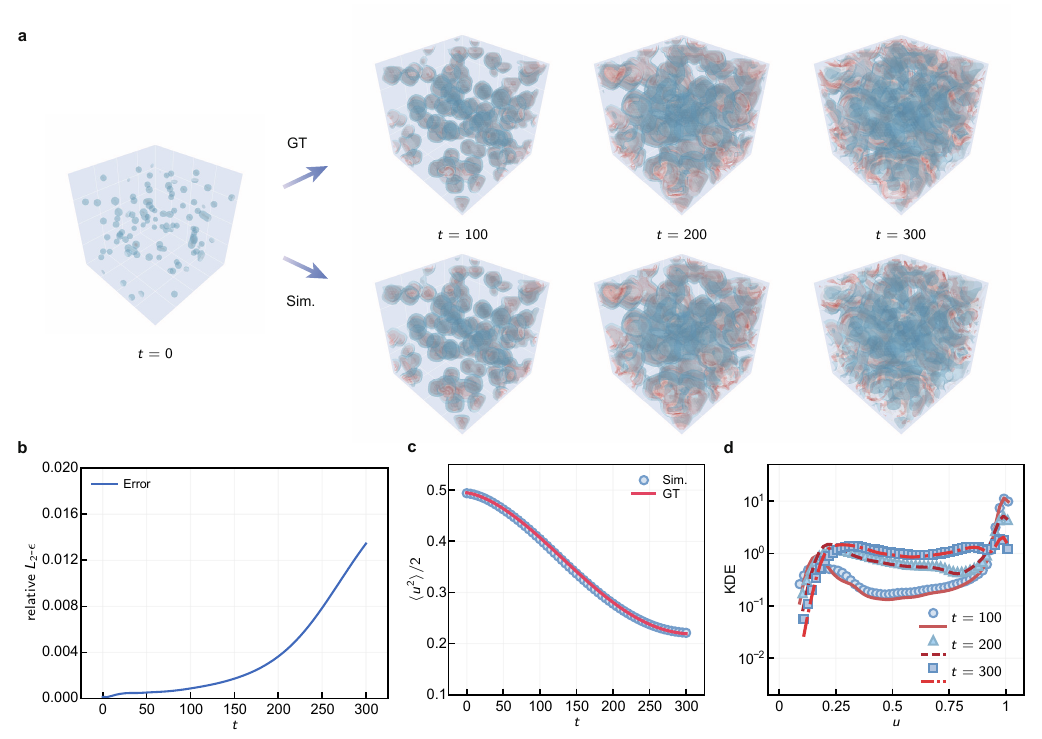}
    \caption{
        Classical emulation of 3D reaction-diffusion systems, comparing with GT.
        (a) Contours of species concentration $u$ at $t=0$ (initial condition, left) and at $t = 100, 200$, and $300$ for GT (top row) and classical emulation.
        (b) Temporal evolution of the relative $L_2$ error.
        (c) Average energy $\langle u^2 \rangle / 2$ as a function of time for QKM (blue circles) and GT (red solid line). 
        (d) KDEs of $u$ at $t = 100$ (circles/solid), $t = 200$ (triangle/dashed), and $t = 300$ (squares/dash--dot), comparing emulation (blue markers) with GT (red lines).
    }
    \label{fig:cube_ideal}
\end{figure*}

The classical emulation of the QKM reproduces the pattern formation of the species concentration $u$. As shown in Fig.~\ref{fig:cube_ideal}(a), the simulated structures match the ground truth (GT) at $t = 100, 200$, and $300$. Quantitative assessment in Fig.~\ref{fig:cube_ideal}(b) shows that the relative $L_2$ error remains below 0.015 throughout the evolution. This is substantially smaller than the error observed during hardware execution ($\sim 0.08$), indicating that the theoretical error $\epsilon_{\text{th}}$ is small and hardware noise $\epsilon_{\text{noise}}$ is the dominant error source in the physical experiment. Furthermore, the average energy $\langle u^2 \rangle / 2$ (Fig.~\ref{fig:cube_ideal}c) and kernel density estimations (Fig.~\ref{fig:cube_ideal}d) closely align with the GT results, confirming that the theoretical framework accurately captures the dissipative dynamics and statistical properties of the system.

\subsection{Spherical fluid dynamics}\label{sec:sim_sphere}

\begin{figure*}
    \centering
    \includegraphics[scale=1.0]{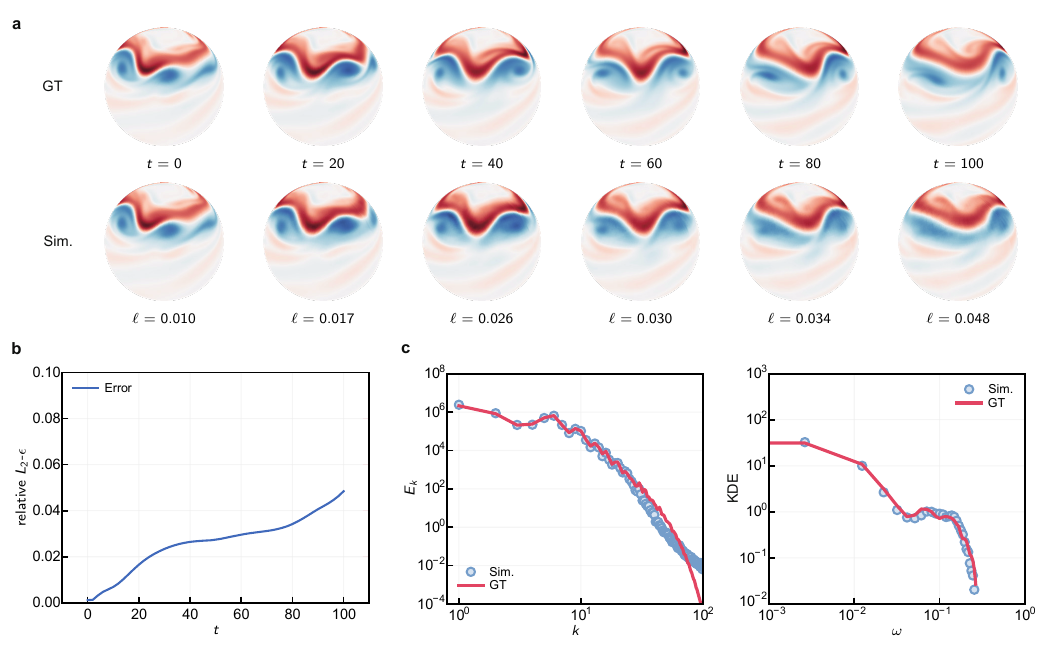}
    \caption{
        Classical emulation of spherical fluid dynamics, comparing with GT.
        (a) Vorticity field $\omega$ at $t=$0, 20, 40, 60, 80, and 100 for GT (top row) and classical emulation, with the relative $L_2$ error $\ell$ indicated beneath each snapshot. 
        (b) Temporal evolution of the relative $L_2$ error. 
        (c) Enstrophy spectrum $E_\omega$ as a function of wavenumber $k$ (left) and KDE of $\omega$ (right), comparing emulation (blue circles) with GT (red solid line) at $t = 80$.
    }
    \label{fig:sphere_ideal}
\end{figure*}

The classical emulation of the QKM reproduces the dynamics of the spherical fluid system. As shown in Fig.~\ref{fig:sphere_ideal}(a), the simulated vorticity field $\omega$ matches the GT from $t = 0$ to $t = 100$. The relative $L_2$ error (Fig.~\ref{fig:sphere_ideal}b) accumulates gradually but remains below 0.06 throughout the evolution. Additionally, the statistical properties at $t = 80$, including the enstrophy spectrum and the kernel density estimation of $\omega$ (Fig.~\ref{fig:sphere_ideal}c), align with the GT, confirming that the theoretical framework preserves the spectral backbone of the system. Comparing this noiseless baseline with the hardware execution error reported in the main text, the theoretical error $\epsilon_{\text{th}}$ and hardware noise $\epsilon_{\text{noise}}$ are of similar magnitude. This indicates that for complex multiscale dynamics, both the theoretical constraints and hardware infidelities contribute comparably to the total simulation error.

\subsection{Real-world ocean currents}\label{sec:sim_square}

\begin{figure*}
    \centering
    \includegraphics[scale=1.0]{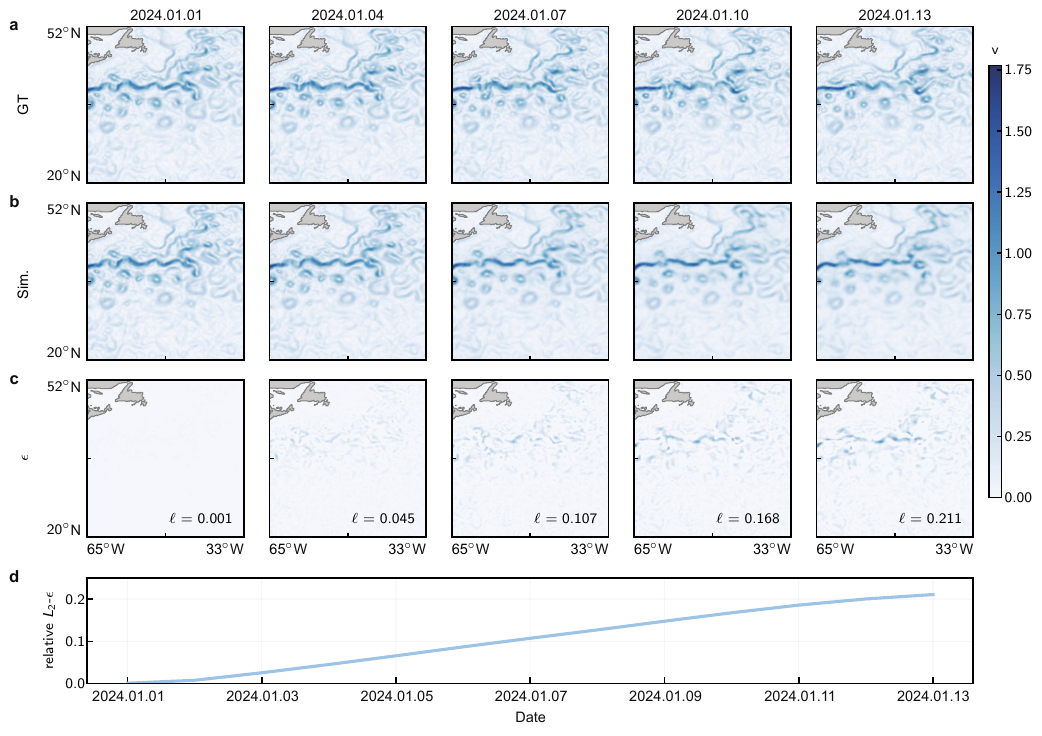}
    \caption{
        Classical emulation of real-world Gulf Stream ocean currents, comparing with the observational data. 
        (a) Surface geostrophic velocity $v$ from satellite-derived GT at five dates spanning January 1-13 2024.
        (b) Corresponding emulation results.
        (c) Pointwise absolute error $\epsilon(x,y;t) = \bigl|v^{\mathrm{Sim}}(x,y;t) - v^{\mathrm{GT}}(x,y;t)\bigr|$, annotated with the relative $L_2$ error $\ell$ indicated beneath each snapshot.
        (d) Relative $L_2$ error as a function of date over the 13-day evaluation period.
    }
    \label{fig:square_ideal}
\end{figure*}

The classical emulation of the QKM reconstructs the mesoscale features of the surface geostrophic velocity field $v$ over a 13-day period. Comparing the satellite-derived ground truth in Fig.~\ref{fig:square_ideal}(a) with the emulation results in Fig.~\ref{fig:square_ideal}(b), the simulated fields align with the observational data. The pointwise absolute error (Fig.~\ref{fig:square_ideal}c) shows that deviations are primarily localized along the high-gradient regions of the current. As shown in Fig.~\ref{fig:square_ideal}(d), the $\varepsilon_{L2}$ exhibits a gradual accumulation typical of chaotic dynamics, reaching approximately 0.21 after 13 days. Because this noiseless baseline error closely matches the total error observed during hardware execution, it indicates that the theoretical error $\epsilon_{\text{th}}$ is large and dominates the overall error budget for this real-world benchmark. 
This systematic bias is primarily attributed to the structural simplification of the evolution operator, where each of the $h$ independent circuits utilizes a single-layer $R_z$ ansatz to approximate the diagonalized spectral components. Improving simulation fidelity for complex geophysical flows thus necessitates refining the ansatz expressivity alongside the finite-dimensional Koopman representation.

\section{Ablation study}\label{sec:ablation}

To substantiate the theoretical scaling laws and error decomposition of the QKM, we perform ablation studies using the spherical fluid dynamics benchmark through classical simulations. The multi-scale interaction and non-trivial topology of this system provide a representative environment to evaluate how the spectral sampling resolution $h$ affects resource efficiency and how the circuit ansatz structure determines the theoretical accuracy floor. In the following, we provide empirical evidence for the $h = \mc{O}(n)$ scaling while demonstrating the dominance of $\epsilon_\text{ansatz}$ within the cumulative theoretical error $\epsilon_\text{th}$.

\subsection{About Theorem~\ref{theory:2}}\label{sec:ablation_h}

We evaluate the impact of the spectral sampling resolution $h$ by testing various values in the vicinity of $n$. According to our error analysis, the selection of $h$ must balance the theoretical spectral sampling error $\epsilon_{\text{spec}}$ established by Theorem~\ref{theory:2} against computational resource overhead. 
Our experimental results demonstrate that the linear scaling $h = \mc{O}(n)$ is sufficient to suppress $\epsilon_{\text{spec}}$ and minimize the cumulative theoretical error $\epsilon_{\text{th}}$ while maintaining a minimal generalization gap. 

\begin{figure*}
    \centering
    \includegraphics[width=\textwidth]{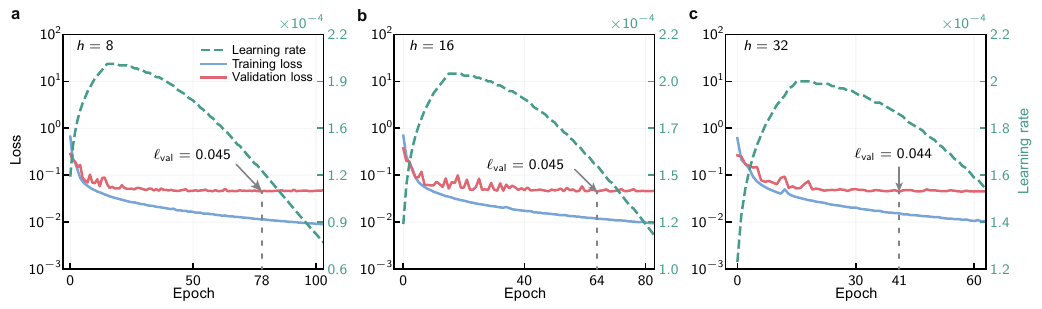}
    \caption{Ablation study of circuit count $h$ for the spherical fluid dynamics benchmark. The system is configured with $(n, R, r) = (10, 3, 3)$. The temporal evolution of the composite loss function $\mc{L}$ and the learning rate schedule are shown for (a) $h = 8$, (b) $h = 16$, and (c) $h = 32$. In each panel, the training loss $\ell_{\text{train}}$ (blue solid line) and validation loss $\ell_{\text{val}}$ (red solid line) are plotted against the left axis on a logarithmic scale. The learning rate schedule (green dashed line), featuring a warm-up phase followed by a cosine decay, is shown on the right axis. The annotated $\ell_{\text{val}}$ indicates the terminal validation error achieved at the conclusion of training via an early stopping criterion.}
    \label{fig:ablation_h}
\end{figure*}

As shown in Fig.~\ref{fig:ablation_h}, increasing $h$ explicitly accelerates the convergence of the training process, with the early stopping criterion triggered progressively earlier at epochs 78, 64, and 41 for $h=8, 16$, and $32$, respectively. However, the terminal validation loss ($\ell_{\text{val}}$) plateaus at approximately 0.044 to 0.045 across all cases, indicating no substantial gain in final predictive accuracy despite the increased circuit ensemble size. These results demonstrate that the linear scaling $h = \mc{O}(n)$ is already sufficient to suppress $\epsilon_{\text{spec}}$ and capture the essential spectral components without incurring exponential resource costs. This empirical evidence supports that the QKM achieves physically meaningful accuracy with a polynomial number of parallel circuits, thereby securing the algorithmic speedup $\mc{S} = \mc{O}(2^n/n^3)$.
Furthermore, as we continue to scale up the framework with a larger $h$, the representational capacity of the model expands. Consequently, larger and more diverse datasets will be required in future large-scale implementations to effectively constrain the optimization landscape. 

\subsection{About Theorem~\ref{theory:3}}\label{sec:ablation_rzz}

We further decompose the theoretical error $\varepsilon_{\text{th}}$ to identify the primary bottleneck in simulation accuracy for multi-scale systems. Theorem~\ref{theory:3} suggests that the structural bias of the single-layer $R_z$ ansatz, $\varepsilon_{\text{ansatz}}$, arises from the high-order Pauli-$Z$ coefficients in the Koopman generator's spectral decomposition. 

\begin{figure*}
    \centering
    \includegraphics[width=\textwidth]{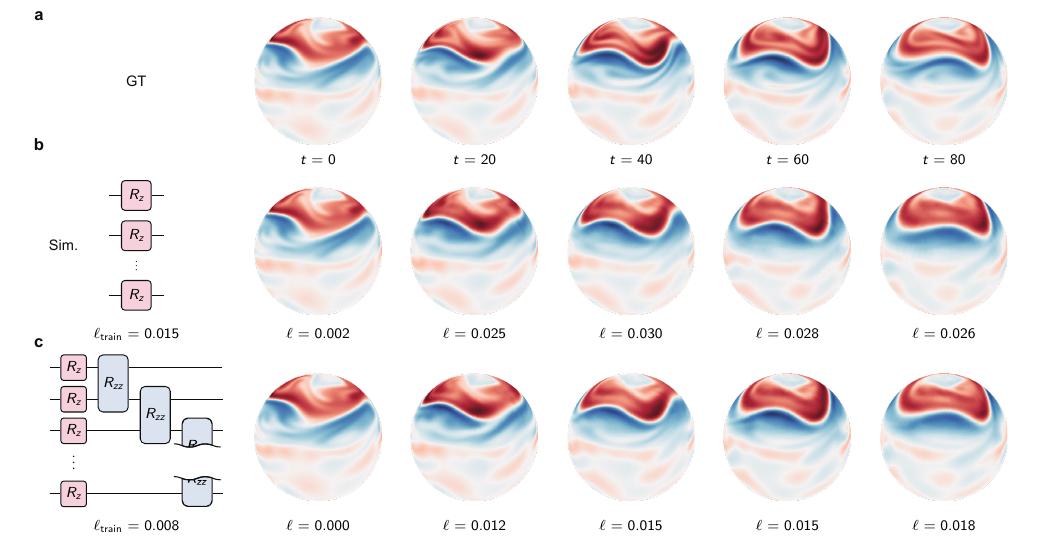}
    \caption{Ablation study on the time-evolution ansatz for the spherical fluid dynamics benchmark. The system is configured with $(n, h, R, r) = (10, 32, 3, 3)$ and the visualized field snapshots are evaluated on samples from the training dataset. 
    (a) GT snapshots of the vorticity field $\omega$ at $t=0, 20, 40, 60$ and $80$.
    (b, c) Classical emulation results with the pointwise relative $L_2$ error $\ell$ indicated beneath each snapshot.
    Specifically, (b) displays the evolution using the baseline single-layer $R_z$ ansatz (first-order approximation), achieving a terminal training loss $\ell_{\text{train}} = 0.015$; (c) displays the evolution using an augmented ansatz that incorporates adjacent two-qubit $R_{zz}$ gate pairs $\{(1,2),(2,3),\ldots,(n,1)\}$ (capturing partial second-order interactions).
    Without altering other hyperparameters, the augmented ansatz further reduces $\ell_{\text{train}}$ to $0.008$, directly demonstrating the mitigation of the structural bias $\epsilon_{\text{ansatz}}$.
    Governed by an early stopping criterion, this reduction represents a genuine suppression of theoretical error rather than overfitting.
    }
    \label{fig:ablation_rzz}
\end{figure*}

To empirically evaluate this, we augment the time-evolution block by introducing $R_{zz}$ gates between adjacent qubit pairs $\{(1,2),(2,3),\ldots,(n,1)\}$, thereby incorporating partial second-order interactions while strictly preserving all other hyperparameters. As illustrated in Fig.~\ref{fig:ablation_rzz}, this augmentation explicitly drives the terminal training loss $\ell_{\text{train}}$, which quantitatively approximates the theoretical error $\varepsilon_{\text{th}}$, down from 0.015 to 0.008. 
The visualized field snapshots are evaluated on the training dataset.
Because the training process is governed by an early stopping criterion, this significant reduction represents a genuine suppression of theoretical error rather than an artifact of overfitting.

This finding confirms that $\varepsilon_{\text{ansatz}}$ is the dominant constituent of $\varepsilon_{\text{th}}$ in our experiments. While the spectral sampling error $\varepsilon_{\text{spec}}$ can be theoretically bounded and systematically suppressed by increasing $h$ according to Theorem~\ref{theory:2}, and the projection error $\varepsilon_{\text{proj}}$ is empirically presumed to be small based on established principles of classical reduced-order modeling, the structural bias of the shallow circuit ultimately governs the theoretical accuracy floor. These results highlight that capturing higher-order mode interactions is the critical pathway for extending the QKM toward more strongly nonlinear regimes.

However, it is worth noting that while the augmented ansatz successfully minimizes $\ell_{\text{train}}$, the validation loss remains largely unchanged. This plateau implies that merely enhancing the expressivity of the quantum circuit is insufficient on its own. To effectively translate the reduced theoretical error into improved generalization and further lower the optimization error $\varepsilon_{\text{opt}}$, it is imperative to scale up the training datasets to adequately constrain the expanded parameter space of the augmented higher-order ansatz.

\bibliographystyle{modified-apsrev4-2.bst}
% \bibliography{supp}
%bst derived from apsrev4-2.bst (revtex)
\providecommand{\noopsort}[1]{}\providecommand{\singleletter}[1]{#1}%

\end{document}